\documentclass[useAMS,usenatbib]{mnras}

\usepackage{amsmath}
\usepackage{amssymb}
\usepackage{graphicx}
\usepackage{dcolumn}
\usepackage{bm}
\usepackage{ulem}

\title[{\it Ab initio} thermodynamics of one-component plasma] {{\it Ab initio} thermodynamics of one-component plasma for astrophysics of white dwarfs and neutron stars}

\author[D. A. Baiko and A. I. Chugunov]{D. A. Baiko\thanks{E-mail:baiko@astro.ioffe.ru} and A. I. Chugunov
\\Ioffe Institute, Politekhnicheskaya 26, 194021 Saint Petersburg, Russia}

\begin{document}
\label{firstpage}
\date{Accepted; Received ; in original form}
\pagerange{\pageref{firstpage}--\pageref{lastpage}} \pubyear{2022}
\maketitle

\begin{abstract}
Using path-integral Monte Carlo (PIMC) simulations, we have calculated  
energy of a crystal composed of atomic nuclei and uniform 
incompressible electron background in the temperature and density 
range, covering fully ionized layers of compact stellar objects, 
white dwarfs and neutron stars, including the high-density regime, where 
ion quantization is important. We have approximated the results by 
convenient analytic formulae, which 
allowed us
to integrate and 
differentiate the energy with respect to temperature and density to 
obtain various thermodynamic functions such as Helmholtz free 
energy, specific heat, pressure, entropy etc. 
In particular, we have demonstrated, that the total crystal specific 
heat can exceed the well-known harmonic lattice contribution by a 
factor of 1.5 due to anharmonic effects. By combining 
our results with the PIMC thermodynamics of a quantum Coulomb liquid, 
updated in the present work, 
we were able to determine density dependences
of such melting parameters as the Coulomb coupling strength at melting, 
latent heat, and a specific heat jump. Our results are necessary 
for realistic modelling of thermal evolution of compact degenerate 
stars.
\end{abstract}


\section{Introduction}
\label{Introd}
Modelling evolution of white dwarfs (WD), in an attempt to understand 
their extremely diverse observational properties, is a hot topic of 
modern astrophysics. Thanks to {\it Gaia} \citep[][]{G16}, the wealth 
of experimental data, awaiting theoretical explanation, is rapidly 
growing. For instance, a population of WD called the Q branch on the 
Hertzsprung--Russell diagram has been recently discovered 
\citep[][]{G18}, which implies an $\sim 8$ Gyr cooling delay with 
respect to a standard evolutionary track \citep*[][]{CCM19}. Such a 
delay may be due to core crystallization accompanied by latent heat 
release and oxygen sedimentation in massive WD \citep[][]{Tetal19}. 
Other sources of extra thermal energy associated with chemical 
separation and gravitational energy release in crystallizing matter 
have been also discussed 
\citep*[e.g.][]{BSBetal20,CHC20,BDS21,CATetal21,BD21,CFHetal21}. 

In this paper, we focus on microphysics of WD interior.
Plasma in a WD core consists of fully ionized atomic nuclei and 
electrons. Its energy has three major terms 
(e.g. \citealt*{HPY07,PC10,B14,OHKetal17}): 
the electron energy (the 
energy of a degenerate electron gas), the ion energy (the energy of an 
ion plasma with constant and uniform electron background), and the 
ion-electron energy (due to electron screening of inter-ion 
interactions).

The electron energy is a well studied quantity. Its 
main term is due to the ideal zero-temperature relativistic degenerate 
electron gas, and there are several higher-order corrections (thermal, 
exchange, correlation). By definition, at a given temperature and 
density, the 
electron contribution is identical regardless of whether ions 
constitute a liquid or a crystal (see below). For many thermodynamic 
quantities, such as pressure and energy, the electron contribution is 
dominant. However, specific heat and temperature derivative of 
pressure are dominated by ions. The latter quantities are crucial for 
thermal evolution and asteroseismology of WD. 

By contrast, the ion-electron thermodynamic functions are not known 
very reliably. They have been studied by two independent methods: 
perturbatively in the liquid \citep[e.g.][]{PC00} and within 
the framework of the harmonic lattice theory in the solid 
\citep[e.g.][]{B02}. As stressed in the latter work, the two 
approaches yielded
contradictory results, predicting, for instance, opposite signs of 
ion-electron corrections to the specific heat in liquid and 
solid phases near melting. This discrepancy remains one of the 
outstanding issues in the field of WD microphysics, however, the 
relative magnitude of these contributions is small, so that, most 
likely, it is inconsequential from the WD evolutionary theory 
perspective.    

In what follows, we shall neglect the higher-order electron and the 
ion-electron contributions, focusing on the ion contribution and 
utilizing the standard thermodynamics of the ideal fully-degenerate 
electron gas only in Sec.\ \ref{Sec_Melt}.

The ion contribution to the thermodynamic quantities is nontrivial. The 
ion plasma with rigid charge-compensating background has been studied 
extensively, as it is a model system for a branch of plasma physics, 
dealing with strongly coupled Coulomb plasmas. If such a plasma 
contains ions of only one sort, it is called a one-component plasma 
(OCP). In general however, one expects a mixture of several ion 
species in WD interior, i.e. a multi-component plasma (for instance, 
a C/O mixture with traces of Ne). Both one- and multi-component 
plasmas have been studied by {\it classic} Monte Carlo and molecular 
dynamics methods (e.g. 
\citealt*{SDWS90,FH93,C99,DS99,DS03,HBB07,HSB10,CHC20,Caplan20,BD21}). 
For instance, the first-order 
melting/crystallization phase transition between a liquid and a 
body-centered cubic solid in a {\it classic} ion OCP is well-known. It 
occurs at $\Gamma = \Gamma_{\rm m} \approx 175$ 
\citep[e.g.][]{PC00,HPY07}, 
where $\Gamma = Z_\mathrm{i}^2 e^2 /(a_\mathrm{i} T)$ is the 
dimensionless Coulomb 
coupling parameter, $T$ is the temperature (the Boltzmann constant 
$k_{\rm B} \equiv 1$), 
$a_\mathrm{i}=(4 \pi n_\mathrm{i}/3)^{-1/3}$ is the Wigner-Seitz 
radius, $n_\mathrm{i}$ and $Z_\mathrm{i}$ 
are the ion number density and charge number, respectively.  

It has been long recognized, that treatment of ions as classic particles 
is not fully adequate in WD interior \citep*[e.g.][]{CADW92}. 
The most obvious illustration of this statement is given by a WD with 
a crystallized core undergoing Debye cooling 
(e.g. \citealt{OA68,LvH75}). To the
lowest
order, the
ion thermodynamics in this case is that of a low temperature Bose gas
of phonons in a harmonic Coulomb solid
\citep*[][hereafter the latter work will be referred to as
Paper I]{K69,C93,BPY01}.
However, anharmonic corrections in the ion crystal are not 
negligible (e.g. \citealt{FH93,PC00}), and, obviously, they are 
also subject to quantum modifications.
Furthermore, for a self-consistent description, 
ion quantum effects should be properly taken into account in 
calculations of the ion liquid thermodynamics as well, at least, at 
not too high temperatures.  

Typically, though,
the ion quantum effects are
included only into the harmonic lattice contribution in the 
solid phase. 
For a few exceptions from this trend, we mention a semi-analytic study 
of quantum melting curve by \citet{C93}, 
based on an extended Lindemann criterion, and a first-principle research 
of \citet*{IOI93} and \citet{JC96}, employing path-integral Monte Carlo 
(PIMC) simulations.   
The latter work, while covering the important physical parameter range
and producing a general picture of ion quantum effects, was not 
detailed enough to allow practical applications of its results to WD 
modelling (see Sec.\ \ref{Sec_Compare} for more details). In spite of
this situation, the ion thermodynamics is tacitly viewed as well-known
in practical astrophysical applications. 

Recently, a new detailed study by the PIMC method of 
quantum ion thermodynamics in the liquid has been carried out 
\citep[][hereafter Paper II]{B19}. In particular, it has been shown, 
that the ion quantum effects
resulted in a sizable reduction of the specific heat at 
temperatures above crystallization. This was especially pronounced for 
heavier WD composed of lighter elements, e.g. helium WD in the 
0.3--0.4 M$_\odot$ mass range or carbon WD exceeding $\sim 1$ M$_\odot$ 
\citep[][hereafter Paper III]{BY19}. Even though these effects were 
obtained for a quantum OCP (quantum multi-component plasmas have not 
been analyzed in detail as of yet), they can be safely expected to 
affect realistic WD and accelerate their cooling. 

Paper III has presented an analytic fit to the energy of the 
quantum one-component ion liquid and has constructed its thermodynamics 
in a closed form. Moreover, the authors analyzed the importance of the 
ion quantum effects for astrophysics of WD, including thermal 
properties, equation of state, and asteroseismologic applications.
In the present paper, we aim to extend these results to the case of 
the quantum one-component ion crystal. In particular, we shall calculate 
the crystal energy from the first principles using the PIMC approach 
(Sec.\ \ref{PIMC}). We shall approximate the energy by an 
analytic formula (Sec.\ \ref{fit_energy}), construct the crystal 
thermodynamics (Sec.\ \ref{fit_thermodyn}) as well as compare our 
results with previous works (Sec.\ \ref{Sec_Compare}). 

An important manifestation of the ion quantum nature is the fact that 
the Coulomb coupling parameter at melting, $\Gamma_{\rm m}$, ceases to 
be a single number and becomes a function of the ion density 
\citep[e.g.][]{C93,JC96}.\footnote{This density dependence of 
$\Gamma_{\rm m}$ due to the ion quantum effects should not be confused 
with a dependence of $\Gamma_{\rm m}$ on the density via the electron 
screening length, which occurs already in a {\it classic} system but 
with a {\it non-rigid} background.}
A combination of our new results in the crystal with those of 
Paper III (updated in Sec.\ \ref{liq_upd}) will allow us to 
re-analyze in detail from the first-principles the OCP properties 
across the phase transition. In particular, we shall establish the exact 
dependence of $\Gamma_{\rm m}$ and the latent heat of crystallization 
on the ion density in the practically relevant range of physical 
conditions, where the ion quantum effects are important 
(Sec.\ \ref{Sec_Melt}). 

In Sec.\ \ref{Sec_apply}, we shall illustrate the ion 
thermodynamics developed in this work by considering a 
few examples of astrophysical relevance.
We conclude in Sec.\ \ref{Sec_conclus} with a summary of the 
most important results.  

To finalize the Introduction, it is worth mentioning, that the physics 
of matter   
in a WD core is essentially the same as that in an 
outer\footnote{It is also applicable to an inner neutron star crust, 
provided that the thermodynamic functions of dripped neutrons can be
separated out.} neutron star crust. 
Thus, our results are fully applicable to these 
fascinating objects as well.

\section{Path Integral Monte Carlo}
\label{PIMC}
%

\begin{figure}                                           
\begin{center}                                              
\leavevmode                                                 
\includegraphics[width=0.47\textwidth,bb=18 8 373 350, clip]
{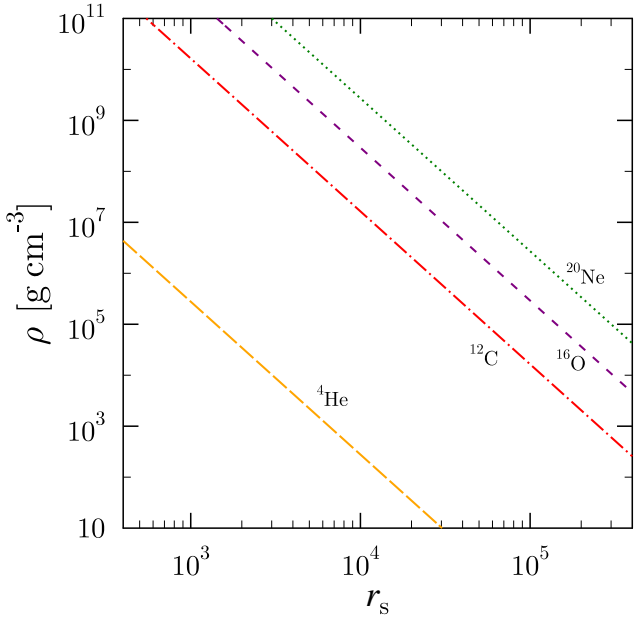} 
\end{center}                                                
\caption{$\rho$--$r_\mathrm{s}$ relationship for several ion species.}                                             
\label{Fig_rs}
\end{figure}
%

We have performed extensive PIMC simulations of body-centered cubic 
(bcc) Coulomb crystals in periodic boundary conditions with $N=250$ 
ions in the density range relevant for astrophysical applications, 
$500 \leq r_\mathrm{s} \leq 120000$, and for temperatures from the 
crystallization, parameterized approximately by the condition 
$\Gamma=175$ (see Sec.\ \ref{Sec_Melt} for update), down to 
$T = T_{\mathrm p}/30$. 
In this case, $r_\mathrm{s} = a_\mathrm{i}/a_{\rm B}$, 
$a_{\rm B} = \hbar^2/(m_\mathrm{i} Z_\mathrm{i}^2 e^2)$ is the ion 
Bohr radius, 
$T_{\mathrm p} = \hbar \sqrt{4 \pi n_\mathrm{i} Z_\mathrm{i}^2 e^2 / 
m_\mathrm{i}}$ is the 
ion plasma temperature, and $m_\mathrm{i}$ is the ion mass. Since 
$r_\mathrm{s}$ does 
not depend on temperature, it determines uniquely density 
$\rho$ for a specified composition. In Fig.\ \ref{Fig_rs}, we show the 
relationship between the mass density and $r_\mathrm{s}$ for several 
nuclei typical of the WD interior.

The general method of calculations followed that described in 
Paper II for the liquid phase of the same system. Specifically, we 
adopt the so-called primitive approximation \citep[e.g.][]{C95}, in 
which the crystal energy, 
\begin{equation}
    \langle {\cal H} \rangle = \frac{ {\rm Tr} \left[{\cal H} 
                              e^{-\beta {\cal H}}\right]}
                    { {\rm Tr} \left[e^{-\beta {\cal H}} \right]}~,
\label{averH}
\end{equation}
where ${\cal H}$ is the system Hamiltonian and $\beta = 1/T$ is the 
imaginary time, can be expressed as
\begin{equation}
        \langle{\cal H} \rangle = \int {\rm d}\sigma \pi(\sigma) 
        H({\cal R}_m)~.
\label{work_eq}
\end{equation}
In this case, $\sigma = \left\{ {\cal R}_1, {\cal R}_2, \ldots, 
{\cal R}_M = {\cal R}_0 \right\}$ is the $3NM$-dimensional domain of 
quantum numbers, i.e. coordinates of $N$ ions in $M$ 
basis states of the coordinate representation. A basis state 
${\cal R}_l$ ($l=1, 2, \ldots, M$) is specified by cartesian coordinates 
of all $N$ ions: ${\cal R}_l = \left\{ {\bm R}^{(l)}_1, {\bm R}^{(l)}_2,
\ldots, {\bm R}^{(l)}_N \right\}$. $M$ is a positive integer, which, in
this formulation, is called the total number of imaginary time slices. 
Equivalently, $\sigma$ is the domain of coordinates of $N$ classic 
ring polymers, each with $M$ beads.   

Furthermore,
\begin{eqnarray}
        \pi(\sigma) &=& \frac{1}{Z} 
        \exp{\left[ -\sum_{l=1}^M S_l \right]}~,
\label{pi(sigma)} \\                                   
           e^{-S_l} &=& \frac{1}{(4 \pi \lambda \tau)^{3N/2}}
      \exp{\left[- 
      \frac{({\cal R}_l-{\cal R}_{l-1})^2}{4 \lambda \tau} 
      - \tau V({\cal R}_l) \right]}~,
\nonumber
\end{eqnarray}
where $Z$ is the partition function, $\lambda = \hbar^2/(2m_{\rm i})$, 
$\tau=\beta/M$, and $V({\cal R})$ is the crystal 
potential energy, when ion coordinates are equal to ${\cal R}$. Finally,
$H({\cal R}_m)$ is the energy estimator. In this work, we use the 
thermodynamic energy estimator given, e.g. by equation (6.7) of 
\citet{C95}. Possible values of $m$, to which the results should be 
insensitive, are $1,2,\ldots, M$.

Since $\pi(\sigma)$ is strictly positive and normalized to 1, it is 
natural to interpret it as a probability distribution, and the task of 
finding $\langle{\cal H} \rangle$ reduces to averaging the energy 
estimator with it. This can be done by sampling with the aid of the 
Metropolis algorithm. We attempt two move types: single bead moves and 
whole polymer moves. 

Unlike in the liquid, the harmonic lattice thermodynamics of the crystal
is well known (Paper I). Hence, it has been decided to separate out the 
anharmonic contribution to the energy. Since anharmonic energy is a 
subdominant contribution (it is smaller than the electrostatic, 
harmonic thermal and harmonic zero-point energies), this required an 
improved precision of numerical computations as compared to the liquid. 
The extra precision was achieved by performing several times longer 
PIMC runs and having multiple 
PIMC runs at fixed physical conditions (i.e.\ temperature and density or
$r_\mathrm{s}$ and $\Gamma$), characterized by different 
random sequences and imaginary time slices $m$, at which 
the energy estimator was applied.  Overall, the typical error bars for 
the present 
anharmonic energy calculations are estimated to be about 5 times 
smaller than the error bars of the liquid energy calculations in 
Paper II. 

The energy estimates from multiple runs at a given temperature, 
density, and a total number $M$ of imaginary time slices (or a number 
of beads in a ring polymer representing a quantum ion) were averaged 
over the runs. Then, the energy of a harmonic crystal was subtracted 
from it. The harmonic crystal energy was calculated for the bcc lattice 
with periodic boundary conditions and the same $N$ and $M$ as in the 
PIMC simulation (see Appendix \ref{AppA} for details). This allowed us 
to prevent 
a contamination of calculated anharmonic energies by relatively large 
finite-size corrections to the harmonic energy.

A detailed numerical analysis at 5 values of $M$ and several 
$(r_\mathrm{s}, \Gamma)$ pairs has shown that the anharmonic energy 
estimates depended on $M$ as 
$c(r_\mathrm{s},\Gamma) - a(r_\mathrm{s},\Gamma)/M^2$, which is the 
same as the $M$-dependence of the harmonic energy 
[cf.\ equation (\ref{ho_correction_to_PIMC})]. 
Having established the quadratic dependence on $1/M$, the data at all 
the other 
physical points were obtained at just two different $M$
and these data were directly used in the fitting procedure.

Since $N$-dependence is not studied quantitatively in this work 
(under assumption that $N=250$ is large enough for this dependence to 
saturate; see also a discussion of $N$-dependence in 
Sec.\ \ref{fit_energy}), the fitted energy is 
treated as the thermodynamic limit of the anharmonic energy. Combined 
with the well-known thermodynamic limit of the harmonic energy 
(Paper I) and with the Madelung energy, it represents our final 
estimate for the ion energy of the crystal.

\section{Analytic expression for the anharmonic crystal energy}
\label{fit_energy}
The ion energy (per ion) reads
\begin{equation}
          U=\frac{\partial F \theta}{\partial \theta}= 
          U_\mathrm{M} + U_\mathrm{h} + U_\mathrm{ah}~,
\label{U}
\end{equation}
where $F$ is the Helmholtz free energy per ion, 
$\theta \equiv T_\mathrm p/T = \Gamma \sqrt{3/r_\mathrm{s}}$, 
indices `M', `h', and `ah' stand for Madelung (static perfect lattice), 
harmonic, and anharmonic 
contributions, respectively; see Paper I for analytic expressions for 
the first and second terms.

The numerical anharmonic contribution to the energy, $U_\mathrm{ah}$, 
can be approximated by a series over anharmonic corrections  
\begin{equation}
     \mathfrak u_{\rm ah} \equiv  
     \frac{U_\mathrm{ah} a_\mathrm{i}}{Z^2 e^2}=
     \frac{A^U_1(\theta)}{\Gamma^2_\mathrm q}
    +\frac{A^U_2(\theta)}{\Gamma^3_\mathrm q}
    +\frac{A^U_3(\theta)}{\Gamma^4_\mathrm q}~,
\label{U_anh}
\end{equation}
where $\Gamma_\mathrm q \equiv Z^2 e^2/(a_\mathrm{i} T_\mathrm p)=
\sqrt{r_\mathrm{s}/3}$ 
and
\begin{eqnarray}
A^U_1(\theta)&=&
   -\frac{A_{11}}{\theta^2}\frac{3A_{12}\theta^2+1}{\left(1+A_{12}\,
   \theta^2\right)^2}
\label{AU1} \\
   &-&\frac{A_{13}}{\theta^2}\frac{3A_{14}\theta^2+1}{\left(1+A_{14}\,
   \theta^2\right)^2} +A_\mathrm{1q}~,
\nonumber \\
A^U_2(\theta)&=&\frac{A_\mathrm{2cl}}{\theta^3}\,
\frac{2-A_{21}\,\theta^4}{2\,\left(1+A_{21}\,\theta^4\right)^{1/4}}~,
\label{AU2} \\
    A^U_3(\theta)&=&\frac{A_\mathrm{3cl}}{\theta^4}+A_\mathrm{3q}.
\label{AU3}    
\end{eqnarray}
In this case, indices `cl' and `q' indicate coefficients, 
determining classic and quantum asymptotes discussed below. 
Moreover, the following set of constraints is assumed to be satisfied 
\begin{eqnarray}
      A_{13} &=& - A_\mathrm{1cl} - A_{11}~,
\label{constr1} \\
      A_{14} A_{13} &=& A_\mathrm{1q} - A_{11}A_{12}~,
\label{constr2} \\
      A_{21}^{3/4} &=& - \frac{2 A_\mathrm{2q}}{A_\mathrm{2cl}}~,
\label{constr3}             
\end{eqnarray}
where the quantities on the left-hand sides are calculated rather than 
fitted. Equations (\ref{constr2}) and (\ref{constr3}) ensure that 
the anharmonic energy has the same, 
$\propto T^4$, 
asymptote as the 
harmonic energy at very low temperatures ($\theta \gg 1$). This 
assumption is unmistakably supported by the PIMC data at 
$\theta \leq 30$. Inclusion of terms of higher-order in anharmonism is 
likely required to extend the fit to extremely high densities, 
$r_\mathrm{s}<500$. This is beyond the scope of the present work, 
because, 
as discussed in Sec.\ \ref{Sec_Melt}, such densities are unrealistic
(see also Fig.\ \ref{Fig_rs}).

In general, the anharmonic energy can be split into two parts, the 
zero-point and the thermal contributions. Both are contained in 
equation (\ref{U_anh}). The thermal part of the anharmonic energy is 
positive. For a classic crystal, only thermal contribution to 
the anharmonic energy exists, and it was studied earlier 
semi-analytically \citep[][]{AG86,D90}, by classic Monte 
Carlo (\citealt{SDWS90}), and by molecular dynamics methods 
\citep[][]{FH93}. According to the results of these studies, the 
classic anharmonic energy can be
presented in the form of an expansion over $\Gamma$ 
\begin{equation}
 \frac{U_\mathrm{ah,cl}}{T} = \frac{A_\mathrm{1cl}}{\Gamma}
  + \frac{A_\mathrm{2cl}}{\Gamma^2} + \frac{A_\mathrm{3cl}}{\Gamma^3}~.
\label{anhexp}
\end{equation}
The functional form of equation (\ref{anhexp}) is 
consistent 
with our data at quasiclassic conditions and is 
reproduced by equation (\ref{U_anh}) in the limit $\theta \to 0$ by 
construction. 

The literature values of the classic anharmonic coefficients in 
equation (\ref{anhexp})
have some scatter \citep[cf.][]{D90,FH93}. In particular, \citet{FH93}
suggest several sets of anharmonic coefficients, depending on 
$\Gamma$ range and the number of anharmonic terms in their fit to the 
numerical data. The 
coefficients vary quite a bit from set to set (for instance, 
$A_\mathrm{1cl}$ varies from 5.98 to 10.9) with little effect on the 
fit accuracy. The set from line 1 of table 5 in this reference:  
$A^{\rm astro}_\mathrm{1cl}=10.9$, 
$A^{\rm astro}_\mathrm{2cl}=247$, and 
$A^{\rm astro}_\mathrm{3cl}=1.765\times 10^5$
is widely employed in astrophysics \citep[e.g.][]{PC10}.
On the other hand, in plasma community 
\citep[e.g.][]{DON99,KK16}, 
the set suggested by \citet{D90} is in frequent use: 
$A^{\rm plasma}_\mathrm{1cl}=10.84$, 
$A^{\rm plasma}_\mathrm{2cl}=352.8$, and 
$A^{\rm plasma}_\mathrm{3cl}=1.794\times 10^5$.

Our PIMC data can be fitted quite satisfactorily with the `cl' fit 
coefficients in equations (\ref{U_anh})--(\ref{constr3}) fixed to the 
``astrophysical'' values above, but if they are fixed to the 
``plasma physics'' values, the fit (\ref{U_anh}) becomes noticeably 
worse (rms error increases by $\sim 40\%$). At the same time, the best 
fit to our data corresponds 
to a bit different set of values: $A_\mathrm{1cl}=10.2$, 
$A_\mathrm{2cl}=248$, and $A_\mathrm{3cl}=2.03\times 10^5$ 
(cf.\ Tab.\ \ref{Table_FitPar}), which lie within the scatter of 
published classic anharmonic coefficients.

\begin{table}
    \begin{tabular}{r|r}
                    \hline\hline
    $A_\mathrm{1cl}$        &     $10.2$\\ 
    $A_\mathrm{2cl}$        &      $248$ \\
    $A_\mathrm{3cl}$        &     $2.03\cdot10^{5}$ \\
    $A_{1\mathrm q}$  &     $-0.62/6$\\ 
    $A_{2\mathrm q}$  &     $-0.56$ \\
    $A_{3\mathrm q}$  &     $2.35$ \\
    $A_{11}$          &      $-10$ \\
    $A_{12}$          &      $6\cdot10^{-3}$ \\
    $B_1$          &      $0.13$ \\
    $B_2$          &      $4\cdot10^{-4}$ \\
    rms err           &   $5\times 10^{-6}$\\
    max err           &   $1.5\times 10^{-5}$ \\
    \hline\hline
    \end{tabular}
    \caption{The fitting parameters}
    \label{Table_FitPar}
\end{table}
%

As shown by \citet*{Carr_ea61}, \citet{C78}, and 
\citet{AG86}, the anharmonic zero-point 
energy can be expanded in powers of $r_\mathrm{s}^{-1/2}$, 
with the lowest-order term being negative and inversely proportional to 
$r_\mathrm{s}^2$. Our numerical data and fit reproduce this behavior 
with a 
slightly different coefficient. Namely, we get 
$6A_\mathrm{1q}=-0.62$, whereas \citet{Carr_ea61} and \citet{C78}  
had, for the same 
quantity, $-0.73$, while \citet{AG86} reported $-0.703$. 
As above, the PIMC data can be fitted satisfactorily
with the values for $6A_\mathrm{1q}$ reported by previous authors, 
whereas $-0.62$ is our best fit value. On top of that, our fit 
expression contains higher-order terms of the anharmonic zero-point 
energy expansion.  

Overall, our fit has 8 numerical parameters in 
equations (\ref{AU1})--(\ref{AU3}), which are listed in 
Tab.\ \ref{Table_FitPar} with minimum required number of digits 
along with the rms and maximum fit errors.
Two additional fitting parameters, $B_1$ and $B_2$, describe the 
dependence of the finite-$M$ correction coefficient 
$a(r_\mathrm{s},\Gamma)$
on $r_\mathrm{s}$ and $\Gamma$ as follows
\begin{equation}
  a(r_\mathrm{s},\Gamma)=3\,B_1\frac{\Gamma^2}{r_\mathrm{s}^2}\,
  \left(1+B_2\,\theta^2\right)^{-1}~.
\end{equation}
This coefficient is not needed for astrophysical applications, but may 
be usefull for subsequent PIMC studies.
 
At intermediate temperatures, where neither classic nor quantum 
asymptotes apply, our results, for the first time, provide the 
detailed first-principle temperature 
and density dependence of the true quantum anharmonic energy and 
thermodynamics of a Coulomb crystal.

\begin{figure}                                           
\includegraphics[width=\columnwidth]
{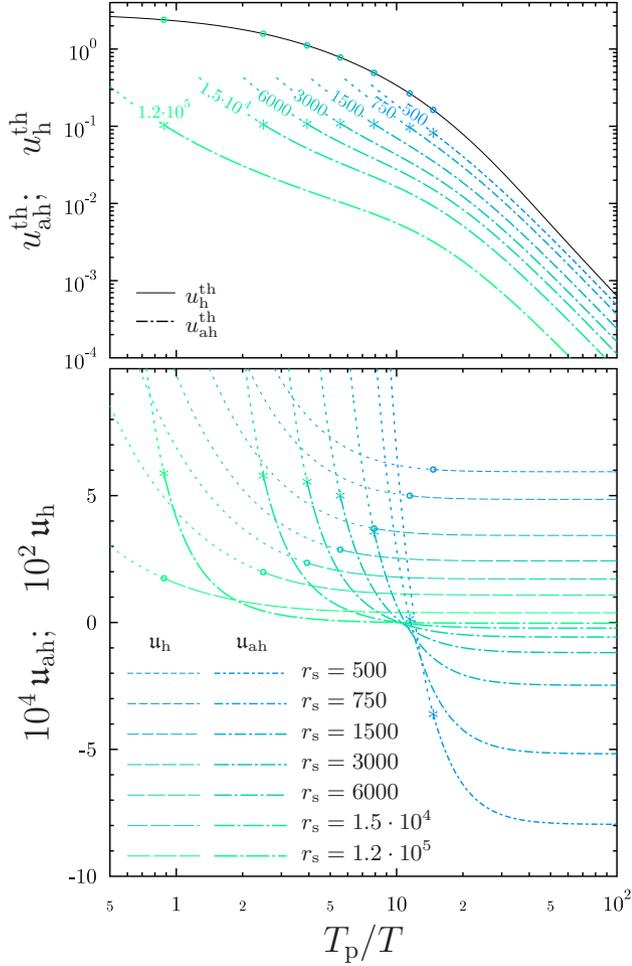} 
\caption{Bottom panel: thin dashed and thick dot-dashed curves show 
harmonic ($\times 10^2$) and anharmonic ($\times 10^4$) contributions 
to the one-component quantum ion crystal energy in units of 
$Z^2 e^2/a_\mathrm{i}$ [$\mathfrak  u_{\rm h, ah} \equiv a_\mathrm{i} 
U_{\rm h, ah}
/(Z^2 e^2)$].
Different density of dashes corresponds to different $r_\mathrm{s}$ 
values. 
Symbols indicate melting temperatures at respective $r_\mathrm{s}$. 
Dotted 
portions of the curves describe a superheated crystal. Top panel: 
thermal harmonic (thin, solid, same for all $r_\mathrm{s}$) and 
anharmonic (thick, dot-dashed) contributions to the energy in  
units of $T$, without multiplication by extra powers of $10$
($u_{\rm h, ah} \equiv U_{\rm h, ah}/T$).
}                                             
\label{Fig_energy}
\end{figure}
%

In Fig.\ \ref{Fig_energy}, bottom panel, we plot harmonic (multiplied by 
$10^2$, thin, dashed) and anharmonic (multiplied by $10^4$, thick, 
dot-dashed) crystal energies in units of $Z^2 e^2/a_\mathrm{i}$ as 
functions of 
$T_{\rm p}/T$ for several values of $r_\mathrm{s}$ in the range 
$500 \leq r_\mathrm{s} \leq 120000$. Symbols indicate melting 
temperatures with 
account of the $\Gamma_{\rm m}(r_\mathrm{s})$ dependence established in 
Sec.\ \ref{Sec_Melt}. At higher temperatures (lower $T_{\rm p}/T$), 
the system is assumed to be in a superheated crystal state (dotted 
portions of the curves). The anharmonic energies are given by the 
present fit, while harmonic energies are taken from Paper I. 

Top panel 
shows only the thermal (`th') harmonic and anharmonic contributions to 
the energy in units of $T$ and without multiplication by extra numerical 
factors. If expressed in units of temperature, harmonic 
contributions to the energy (as well as to the Helmholtz free energy, 
cf.\ Fig.\ref{Fig_helmholtz}) are independent of $r_\mathrm{s}$.   

Since the total anharmonic energy is a sum of the (negative) zero-point 
and (positive) thermal contributions, at any density, there is a 
temperature where the total anharmonic energy is zero. For 
$r_\mathrm{s}>600$, 
it takes place at $T=(10-12)T_\mathrm p$.
Even more 
peculiar is the fact that there exists a density 
($r_\mathrm{s}\approx 735$), 
at which the total anharmonic energy is zero at melting! 

Let us note in passing, that for $N=250$ particles, \citet{D90} 
has found analytically $A_\mathrm{1cl}=10.036$, which is closer to 
our best fit value. This represents an 8\% deviation from the 
(analytic) thermodynamic limit
reported by
the same author, $A^{\rm plasma}_\mathrm{1cl}=10.84$, 
due to a finite-$N$ effect at $N=250$. Strong $N$-dependence of the 
classic thermal anharmonic energy has been seen by 
\citet{AG86}, 
however, these authors have also observed unwelcome dependence on 
the mesh type and oscillating $N$-dependence (cf.\ their Figs.\ 1 
and 2). 

In general, 8\% seems to be a surprisingly large finite-$N$ 
correction to an energy at $N=250$.
Naively, one expects the correction to be $\sim 1/N \sim 0.4\%$.
In fact, this is what we find for 
the finite-$N$ correction to the harmonic zero-point energy and to the 
harmonic classic entropy, determined by the phonon frequency moments
$\langle \omega/\omega_{\rm p} \rangle$ and 
$\langle \ln{(\omega/\omega_{\rm p})} \rangle$, respectively
(see Fig.\ \ref{Fig_FreqMom} in the Appendix \ref{AppA}; 
$\omega_{\rm p} = T_{\rm p}/\hbar$ is the ion plasma frequency). 
Classic thermal harmonic 
energy has $N$-correction of the same, $1/N$, 
order [e.g. equation (18) of \citet{D90}]. Note also an $\approx 1/N$
difference between our best fit $A_\mathrm{2cl}$ and 
$A^{\rm astro}_\mathrm{2cl}$, even though the latter coefficient is 
obtained from a simulation with $N=1024$ particles. Finally, Fig.\ 2 
of \citet{AG86} seems to indicate that the lowest-order 
anharmonic zero-point energy does not demonstrate strong 
$N$-dependence either.

On the other hand, the formulae derived by \citet{D90} for 
$A_\mathrm{1cl}$ have phonon frequencies in the denominator, whereas 
the phonon frequency moment 
$\langle \omega_{\rm p}^2 /\omega^2 \rangle$ does have an 
$N$-correction of a similar order (cf.\ Fig.\ \ref{Fig_FreqMom}). 
Thus, summarizing the discussion, we should warn the reader that our fit 
for the anharmonic energy may be affected by finite-size effects 
especially noticeable (at the level of $\sim 8\%$) in the lowest-order 
anharmonic classic term. It would require significantly more computer 
resources to get rid of said effects in a first-principle PIMC 
simulation.

\section{Thermodynamics of the crystal}
\label{fit_thermodyn}
The ion Helmholtz free energy (per ion) can be written as
\begin{equation}
    F=F_\mathrm{M} +F_\mathrm{h}+F_\mathrm{ah}~,
\end{equation}
where the Madelung, harmonic, and anharmonic terms can be obtained by 
thermodynamic integration from the respective terms in equation 
(\ref{U}). 
In particular, the anharmonic term also takes the form of a series 
over anharmonic corrections
\begin{equation}
   \mathfrak f_{\rm ah} \equiv  \frac{F_{\rm ah} a_\mathrm{i}}{Z^2 e^2}=
        \frac{A^F_1(\theta)}{\Gamma^2_\mathrm{q}}
          +\frac{A^F_2(\theta)}{\Gamma^3_\mathrm{q}}
          +\frac{A^F_3(\theta)}{\Gamma^4_\mathrm{q}}~.
\label{F_anh}
\end{equation}
The 
respective
functions 
are
\begin{eqnarray}
    A^F_1(\theta)&=&\frac{A_{11}}{\theta^2}\frac{1}{1+A_{12}\,\theta^2}
        +\frac{A_{13}}{\theta^2}\frac{1}{1+A_{14}\,\theta^2}
        +A_\mathrm{1q}~,
\label{AF1} \\
    A^F_2(\theta)&=&-\frac{A_\mathrm{2cl}}{2 \theta^3}\,
    \left(1+A_{21}\theta^4\right)^{3/4}~, 
\label{AF2} \\
    A^F_3(\theta)&=&-\frac{A_\mathrm{3cl}}{3 \theta^4}+A_\mathrm{3q}~.
\label{AF3}
\end{eqnarray}

Once the free energy is known, it is straightforward to derive all
the other thermodynamic functions. For instance, the ion entropy $S$ 
(per ion) is
\begin{equation}
S=-\left. \frac{\partial F}{\partial T}\right|_{V,N}
= \frac{\theta^2}{T_{\rm p}}\,
 \frac{\partial F }{\partial \theta}
=\frac{U-F}{T}
=S_\mathrm{h}+S_\mathrm{ah},
\end{equation}
where 
\begin{equation}
S_\mathrm{ah}= \frac{\theta^2}{T_{\rm p}} \,
\frac{\partial F_\mathrm{ah}}{\partial \theta}=
\frac{A^S_1(\theta)}{\Gamma_\mathrm{q}}
+\frac{A^S_2(\theta)}{\Gamma^2_\mathrm{q}}
+\frac{A^S_3(\theta)}{\Gamma^3_\mathrm{q}}.
\label{S_anh}
\end{equation}
The analytic form of the coefficients is:
\begin{eqnarray}
A^S_1(\theta)&=&
-\frac{2 A_{11}}{\theta}
\frac{2\,A_{12}\theta^2+1}{\left(1+A_{12}\,\theta^2\right)^2}
\nonumber \\ &-&
\frac{2 A_{13}}{\theta}
\frac{2\,A_{14}\theta^2+1}{\left(1+A_{14}\,\theta^2\right)^2}~,
\\
A^S_2(\theta)&=&\frac{3\,A_\mathrm{2cl}}{2\,\theta^2}
\frac{1}{\left(1+A_{21}\,\theta^4\right)^{1/4}}~,
\\
A^S_3(\theta)&=&\frac{4\,A_\mathrm{3cl}}{3\theta^3}~.
\end{eqnarray}

Similarly, the ion specific heat $C$ (per ion) reads
\begin{equation}
C=\left. \frac{\partial S}{\partial \ln T}\right|_{V,N}
= -\theta\frac{\partial S }{\partial \theta}
=C_\mathrm{h}+C_\mathrm{ah},
\end{equation}
where 
\begin{equation}
C_\mathrm{ah}=-\theta
\frac{\partial S_\mathrm{ah} }{\partial \theta}=
\frac{A^C_1(\theta)}{\Gamma_\mathrm{q}}
+\frac{A^C_2(\theta)}{\Gamma^2_\mathrm{q}}
+\frac{A^C_3(\theta)}{\Gamma^3_\mathrm{q}}.
\label{C_anh}
\end{equation}
The analytic form of the coefficients is:
\begin{eqnarray}
A^C_1(\theta)&=&
  -\frac{2\,A_{11}}{\theta}\,
   \frac{6\,A_{12}^2\theta^4+3\,A_{12}\,\theta^2+1}
        {\left(1+A_{12}\theta^2\right)^3}
        \\
  &-&\frac{2\,A_{13}}{\theta}\,
\frac{6\,A_{14}^2\theta^4+3\,A_{14}\,\theta^2+1}
{\left(1+A_{14}\theta^2\right)^3}~,
       \nonumber
\\
A^C_2(\theta)&=&\frac{3\,A_\mathrm{2cl}}{2\,\theta^2}
\frac{2+3\,A_{21}\theta^4}
{\left(1+A_{21}\,\theta^4\right)^{5/4}}~,
\\
A^C_3(\theta)&=&4\,\frac{A_\mathrm{3cl}}{\theta^3}~.
\end{eqnarray}

Finally, analytic formulae for the ion pressure read: 
\begin{eqnarray}
  P &=& - N \left( \frac{\partial F}{\partial V} \right)_T = 
     P_\mathrm{M}+   P_{\rm h} + P_{\rm ah}~,
\nonumber \\
         \frac{P_{\rm ah} a_\mathrm{i}}{n_\mathrm{i} Z^2 e^2} &=&          
        \frac{S_{\rm ah}}{2 \Gamma} 
        + \frac{2 A^F_1(\theta)}{3 \Gamma^2_\mathrm{q}}
        + \frac{5 A^F_2(\theta)}{6 \Gamma^3_\mathrm{q}}
        + \frac{A^F_3(\theta)}{\Gamma^4_\mathrm{q}}~.
\label{Press}                
\end{eqnarray}
$P_\mathrm{M}=n_\mathrm{i} U_\mathrm{M} /3$ is the (negative) 
electrostatic or Madelung contribution to the pressure. 

Let us remind, that the thermodynamic functions $F_\mathrm{h}$, 
$S_\mathrm{h}$,  $C_\mathrm{h}$, and  $P_\mathrm{h}$  can be easily 
derived from the harmonic lattice thermodynamics fits of Paper I.

\begin{figure}                                           
\begin{center}                                              
\leavevmode                                                 
\includegraphics[width=\columnwidth]
{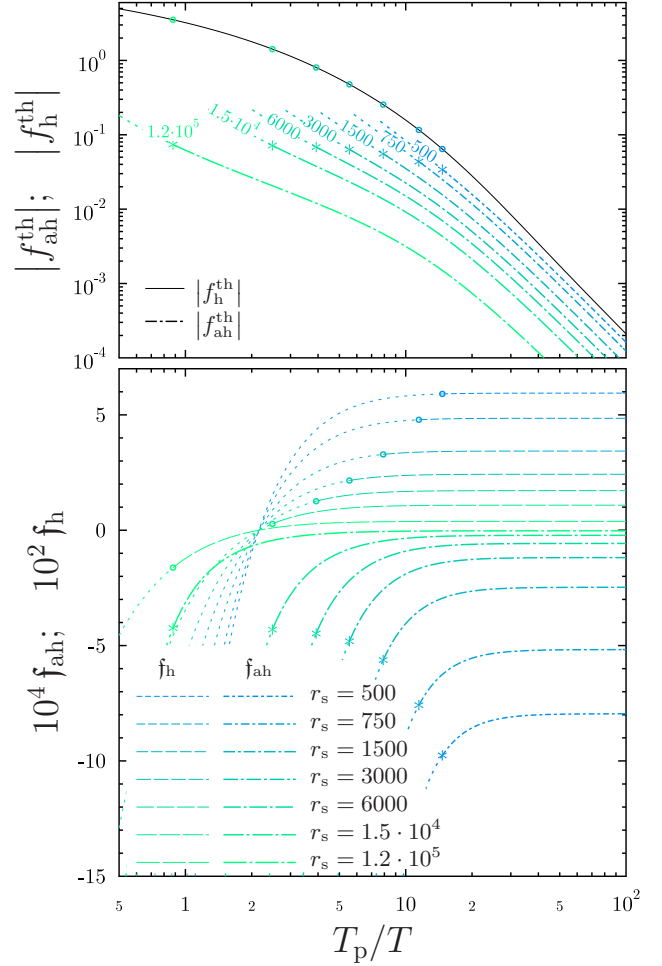}
\end{center}                                                
\caption{Same as in Fig.\ \ref{Fig_energy} but for the crystal ion 
Helmholtz free energy 
[$\mathfrak f_{\rm h, ah} \equiv a_\mathrm{i} F_{\rm h, ah}/(Z^2 e^2)$,
$f_{\rm h, ah} \equiv F_{\rm h, ah}/T$]. 
Top panel shows absolute values of (negative) thermal contribultions 
to the free energy.}
\label{Fig_helmholtz}
\end{figure}

\begin{figure}                                           
\begin{center}                                              
\leavevmode                                                 
\includegraphics[width=\columnwidth]
{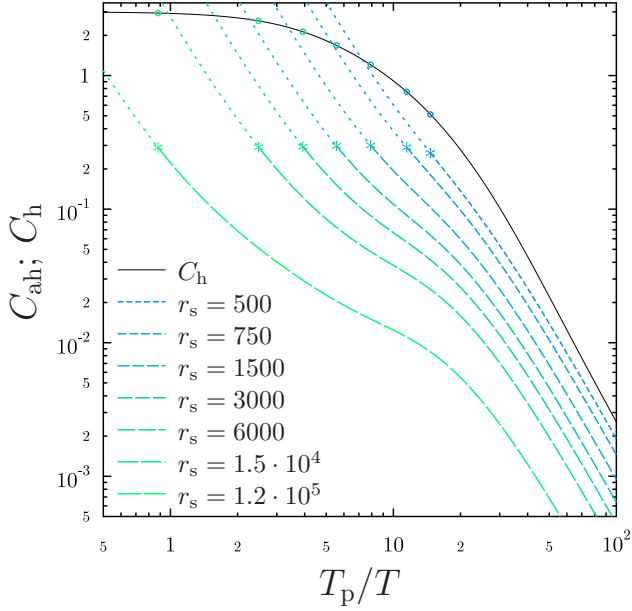} 
\end{center}                                                
\vspace{-0.4cm} 
\caption{Harmonic (thin, solid, one curve for all $r_\mathrm{s}$) and 
anharmonic (thick, dashed) contributions to the specific 
heat of the one-component quantum ion crystal. Symbols indicate melting 
temperatures. Dotted portions of the curves assume a superheated 
crystal. 
}                                             
\label{Fig_heat}
\end{figure}

In Figs.\ \ref{Fig_helmholtz} and \ref{Fig_heat}, we show crystal 
ion Helmholtz free-energy and ion specific heat as functions of 
$T_{\rm p}/T$ for several $r_\mathrm{s}$ values. Thin dashed curves in 
the bottom panel of Fig.\ \ref{Fig_helmholtz} 
display harmonic contributions. Thick dot-dashed curves are anharmonic 
quantities based 
on the present fit. Top panel shows absolute values of the thermal 
contributions to the free energy, which are actually negative. The 
harmonic contribution, drawn by a thin solid curve, is insensitive to 
$r_\mathrm{s}$. The Helmholtz free energy will be 
discussed further in Sec.\ \ref{Sec_Melt}.

%

In Fig.\ \ref{Fig_heat}, the harmonic contribution to the ion specific 
heat is shown by a thin solid line, and it is also the same for all 
$r_\mathrm{s}$.
In the quantum limit of low temperatures, the anharmonic 
specific heat (thick, dot-dashed) exhibits the same $T^3$ dependence 
as predicted by the Debye law for the harmonic contribution. At high 
temperatures, the temperature dependence of the anharmonic specific 
heat is more complex. It does not saturate and 
increases rather rapidly with temperature increase, even 
exceeding the harmonic contribution in the superheated phase. 

In the normal crystal phase, the anharmonic specific heat is always 
smaller than the harmonic one, but its relative importance grows, as 
the system becomes denser ($r_\mathrm{s}$ decreases). For instance, at 
$r_\mathrm{s}=120000$, the anharmonic specific heat is about 30 times 
smaller 
than the harmonic one in the quantum asymptote regime, becomes almost 
100 times smaller at $\theta \sim 5$, and is only 10 times smaller at 
melting. By contrast, at $r_\mathrm{s}=600$, the anharmonic 
contribution is 
only about two times smaller than the harmonic one in the $\propto T^3$ 
regime, and the ratio reaches maximum of about 3 at 
$T_\mathrm p \sim 20 T$.        

It is worth noting, that the anharmonic specific heat at the melting 
point is $\sim 0.3$, being almost independent of $r_\mathrm{s}$.

%

\section{Comparison with earlier work}
\label{Sec_Compare}
Let us compare the first-principle quantum thermodynamics of a
one-component ion crystal developed above with several earlier works. 

First-principle PIMC calculations of the Coulomb crystal 
energy have been performed earlier by \citet{IOI93} and \citet{JC96}. 
A comparison of the anharmonic energy, reported by \citet{IOI93} and 
shown in their Fig.\ 6, reveals an agreement with our fit at nearly 
classic temperatures $\theta \lesssim 4$. Deeper into the 
quantum domain, their points deviate from our fit
at all $\Gamma$, and the discrepancy steadily grows with increase 
of $\Gamma$. In the ultra-quantum regime, 
$50 \lesssim r_\mathrm{s} \lesssim 500$, dominated by the zero-point 
asymptote, our fit and their data points are again in a rough 
agreement with each other except at $\Gamma=1000$, where their data is 
poor. Unfortunately, we do not have a way of identifying the source of 
the disagreement at intermediate $\theta$. What is clear is that our 
approach employs more particles, MC iterations, beads, contains an 
accurate subtraction of the harmonic energy at a finite $N$, and does 
not use any analogs of a cumulant expansion. Besides, their prediction 
for the anharmonic entropy \citep[Fig.\ 7 of][]{IOI93} looks 
implausible with a value, 
greatly exceeding the harmonic 
contribution at $T \ll T_\mathrm p$, and a minimum at intermediate 
$\theta$.

\citet{JC96} have presented PIMC calculations with $N=54$ ions of the 
bcc solid phase energy (harmonic plus anharmonic) at 
$r_\mathrm{s}=200$, $400$, $625$, and $1200$. 
We assume that the data points shown in their Fig.\ 2 include the 
finite-$N$ correction described by their equation (1). These 
data points do 
not agree with our anharmonic fit combined with the harmonic energy 
taken in the thermodynamic limit (from Paper I). However, if one removes
the finite-$N$ correction from the data points in Fig.\ 2 of 
\citet{JC96}, the raw 
PIMC data obtained in this way agree reasonably well with our 
anharmonic fit combined with the harmonic energy of a lattice with   
$N=54$ ions (see Sec.\ \ref{Sec_HL_PIMC} for details). We thus conclude
that the data in Fig.\ 2 of \citet{JC96} is likely compromised by 
inaccurate treatment of finite-size effects in a harmonic lattice. 

\begin{figure}                                           
\begin{center}                                              
\leavevmode                                                 
\includegraphics[width=\columnwidth]
{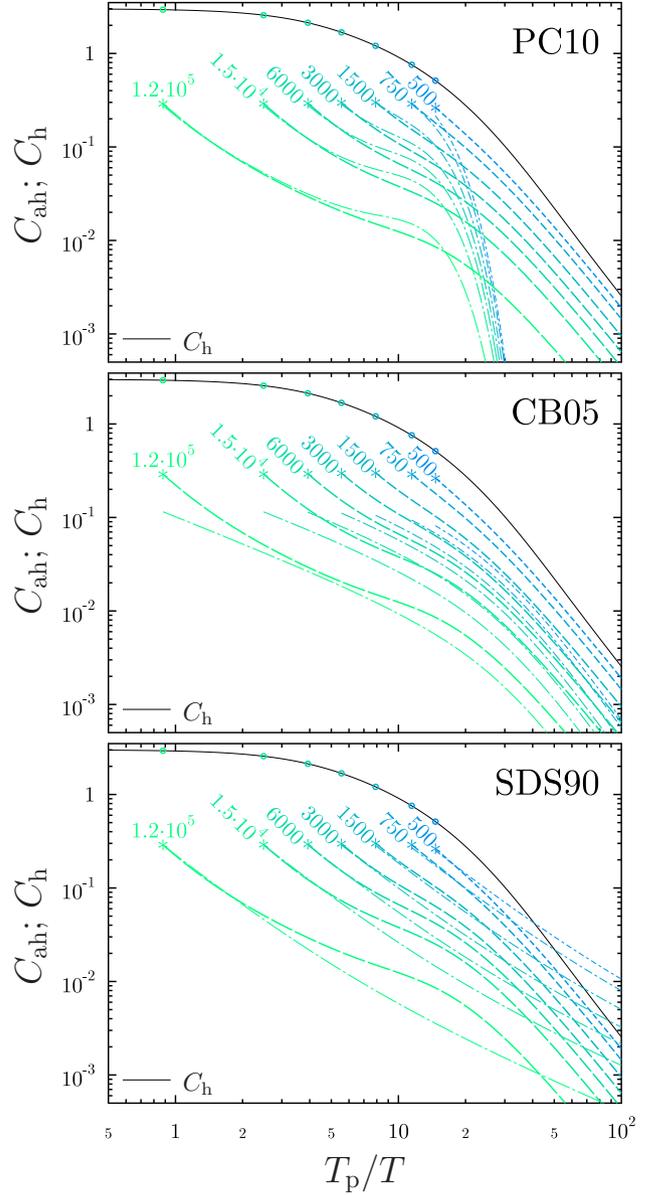} 
\end{center}                                                
\vspace{-0.4cm} 
\caption{Various results for ion crystal specific heat. Thin solid 
lines display the harmonic contribution $C_\mathrm{h}$ from Paper I 
(same for all $r_\mathrm{s}$). Dashes show the anharmonic contributions
calculated in this work (and already shown in Fig.\ \ref{Fig_heat}) at 
$r_\mathrm{s}$ values indicated near the curves. Thin dot-dashed lines
represent previously available parametrizations: \citet{PC10} 
(top panel), \citet{CB05} (middle panel), and \citet{SDWS90} (bottom 
panel).}                                             
\label{fig_sh_compar}
\end{figure}
%

Thus, the most obvious advantage of our calculations over those of 
\citet{IOI93} and \citet{JC96} is their very detailed nature and 
superior accuracy, which allowed us to develop a highly constrained fit.  
The fit can be integrated and differentiated to produce reliable 
thermodynamics free of unphysical artefacts in the astrophysically 
meaningful density range.  

In Fig.\ \ref{fig_sh_compar}, using the specific heat as an example, 
we illustrate the 
difference between our PIMC calculations of the ion crystal 
thermodynamics and several earlier parametrizations available in the 
literature. For reference, by thin solid line, we plot the harmonic
specific heat from Paper I (same as in Fig.\ \ref{Fig_heat}). This 
contribution is a universal function of $\theta$, i.e. it is the same 
for all densities at fixed $\theta$. Dashes (same as in 
Fig.\ \ref{Fig_heat}) show the anharmonic contribution to the specific
heat calculated in this work at the $r_\mathrm{s}$ values indicated near
the curves.    

Thin dot-dashed lines in the top panel display widely used 
interpolation of the anharmonic contribution to the specific heat 
suggested by \cite{PC10}. It assumes an exponential suppression of the 
thermal anharmonic energy and specific heat in quantum regime, 
which clearly disagrees with our data and fit. 

Furthermore, for anharmonic correction to the zero-point energy, this 
interpolation uses only the lowest-order anharmonic term calculated by 
\citet{Carr_ea61}, \citet{C78}, and \citet{AG86}. According to our 
simulations, 
this is not accurate enough at $r_\mathrm{s} \lesssim 1500$.

%
\begin{table*}
\caption[]{PIMC energies in the liquid phase at $r_\mathrm{s}=500$}
\begin{tabular}{rrrrrrrrrr}
\hline      
\hline      
 $\Gamma$ & 90 & 101 & 112 & 123 & 134 & 145 & 156 & 167 & 178 \\
\hline      
  $(U-U_{\rm M})/T$ & 6.344 & 6.951 & 7.565 & 8.185 & 8.811 & 9.438 & 10.072 & 10.707 & 11.343 \\
\hline      
\hline      
\end{tabular}
\label{liq500en}
\end{table*}
%

\citet{CB05} have analyzed quantum anharmonic corrections to the Coulomb 
crystal thermodynamics based on the harmonic pair correlation 
function. The main aim of that paper was to obtain an estimate of the 
quantum anharmonic correction to the crystal specific heat. The 
lowest-order anharmonic correction at arbitrary strength of quantum 
effects was formally included, but the very approach was manifestly 
approximate. For instance, the zero-point properties, obtained in 
\citet{CB05}, were completely inaccurate. Nevertheless, the anharmonic 
crystal specific heat, shown by thin dot-dashed curves in the middle 
panel of Fig.\ \ref{fig_sh_compar}, is seen to be in a rough 
qualitative agreement with the first-principle (dashed) curves.

Thin dot-dashed lines in the bottom panel of Fig.\ \ref{fig_sh_compar} 
represent the specific heat, calculated by differentiation of the 
anharmonic energy given by fits in \citet{SDWS90}. This approach seems 
to be still applied as microphysics input in WD cooling theory 
\citep*[e.g.][]{Segretain_ea94,Salaris_ea21}, even though it completely 
ignores the ion quantum effects for the anharmonic term. These effects 
were not considered by 
\cite{SDWS90}, whose work was based on purely classic Monte Carlo. As 
a result, the anharmonic specific heat is strongly overestimated at low 
temperatures ($T \ll T_\mathrm{p}$), has a wrong asymptote, 
and can even exceed the harmonic contribution, thus corrupting  
modelling of the Debye cooling stage.

%
\begin{table}
\caption[]{Numerical results for $(U-U_{\rm M})/T$ in the liquid phase
near melting}
\begin{tabular}{rrrrrrr}
\hline      
\hline      
              & \multicolumn{6}{c}{$r_{\rm s}$} \\
                    \cline{2-7}
     $\Gamma$~ & 500 & 600 & 750 & 950 & 1200 & 1500  \\
\hline      
  178 & 11.343 & 10.508 &  9.593 &  8.736 &  7.995 &  7.373  \\
  181 & 11.517 & 10.668 &  9.733 &  8.860 &  8.104 &  7.472  \\
  184 & 11.692 & 10.827 &  9.874 &  8.985 &        &         \\
  187 & 11.866 & 10.984 & 10.015 &        &  8.324 &         \\
  190 & 12.040 & 11.143 & 10.156 &        &        &         \\
\hline      
\hline      
\end{tabular}
\label{liq_melt_en}
\end{table}
%

Finally, let us mention that \citet{HV75} have analyzed the 
Wigner-Kirkwood expansion of the free 
energy of the OCP up to the terms of order $\hbar^4$.  
The lowest-order, $\propto \hbar^2$, correction does not depend on the 
OCP state (liquid or crystal). In the solid phase, it is a part of the 
harmonic energy (e.g. Paper I). Moreover, the leading $\hbar^4$-term 
in equation (7) of \citet{HV75} (which is independent of $\Gamma$) 
almost coincides (the relative difference is just $\sim 10^{-3}$)
with the respective term in the Taylor series 
for the harmonic energy. To verify this, we have used the harmonic 
energy fit from Paper I. 

Thus, the remaining terms in equation (7) of \citet{HV75} 
($\propto 1/\Gamma$ and $\propto 1/\Gamma^2$) must correspond to 
quantum corrections to the anharmonic energy. We have compared these 
terms with the Taylor expansion of our anharmonic coefficients, 
equations (\ref{AU1}) and (\ref{AU2}), at $\theta\rightarrow 0$, and 
have found a disagreement by a factor of a few. The most likely reason 
for the discrepancy is the replacement by \citet{HV75} of the three-body 
correlation function by a static-lattice value in their final 
formula for the $K$-term (see \citealt{HV75} for details). Such 
a replacement removes any $\Gamma$-dependence 
from the $K$-term and can make a quantum correction to the anharmonic 
energy inaccurate. 

Note, that already at $\theta \sim 1$, the next 
order, $\propto \hbar^6$, quantum corrections are required, which 
renders the entire WK expansion not very useful \citep[cf.\ a similar 
situation in the liquid as discussed in][and in Paper II]{JC96}.

\section{Liquid phase thermodynamics update}
\label{liq_upd}
%

\begin{figure*}                                           
	\begin{center}                                              
		\leavevmode                                                 
		\includegraphics[width=\textwidth]
		{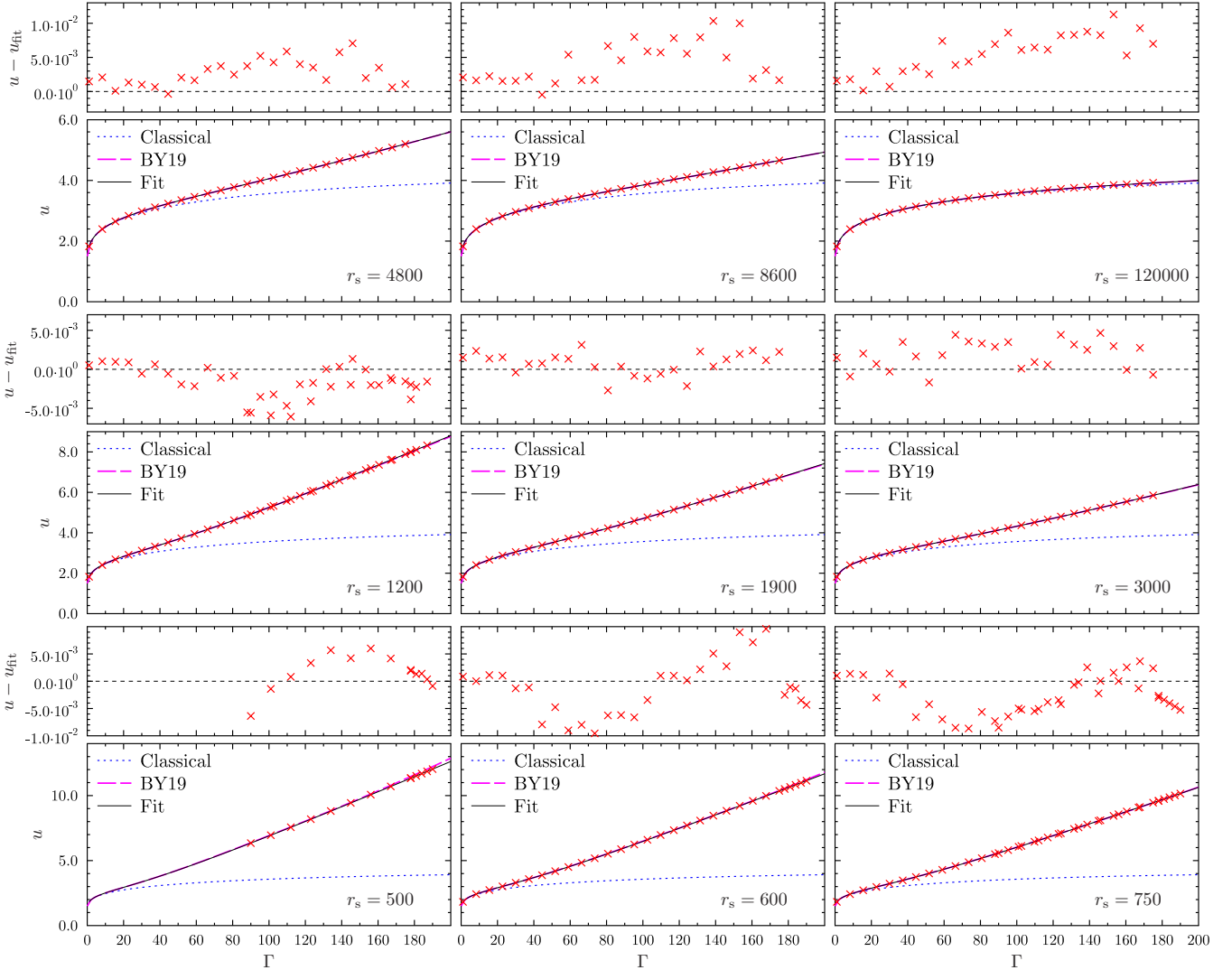} 
	\end{center}                                                
	\vspace{-0.4cm} 
	\caption{PIMC energies in the liquid vs. $\Gamma$ (crosses) for several 
		values of $r_\mathrm{s}$. Thin solid lines represent the fit 
		(\ref{fit_en}) with $u_{\rm q}$ given by equation (\ref{uq123-}), thick 
		dashed lines are the Paper III fit, dotted lines display the 
		right-hand side of equation (\ref{fit_en}) with $u_{\rm q}$ set to 0.}                                             
	\label{Fig_accuracy}
\end{figure*}
%

Once the crystal thermodynamics is constructed, it can be combined with 
liquid thermodynamics of Paper III to obtain a phase diagram of the
strongly-coupled ion plasma at physically relevant densities
and temperatures. Before doing this, we can take advantage 
of the fact, that calculations of the liquid energies were extended by 
one of us to $r_\mathrm{s} = 500$ \citep[][these energies are reported 
here in table \ref{liq500en}]{B21}. Besides that, we have performed 
new PIMC simulations in the liquid phase at the lower end of the 
$r_\mathrm{s}$ range, studied in this paper, in the vicinity of the 
crystallization, which occurs at $\Gamma > 175$ 
(cf.\ Sec.\ \ref{Introd}). These energies are presented in table 
\ref{liq_melt_en}.    

\begin{table}
	\caption{Fit coefficients in equation (\ref{C-coef})}
	\renewcommand{\arraystretch}{1.4}
	\begin{tabular}{cccc}
		\hline      
		\hline      
		 $P_1$ & $P_2$ & $P_3$  \\
		\hline
		 $0.351$   & $0.294$    & $90$ \\
		\hline
		\hline      
	\end{tabular}
	\\
	\label{P-coef}
\end{table}
%

Even though these new data points were in an acceptable agreement 
with the fit proposed in 
Paper III, it was clear that the numerical energies and the fit 
started diverging. This could have resulted in noticeable inaccuracies
of thermodynamic quantities, especially those obtained by fit 
differentiation, in the vicinity of the phase transition and at 
$r_\mathrm{s} \sim 500$. We have therefore modified the fit of 
Paper III to improve the agreement with the new data in the liquid. 
[In addition, it has been noticed by \citet{Jermin_ea21_Skye} that 
the fit of Paper III demonstrated unphysical behaviour in the 
practically unattainable region of very high densities, 
$r_\mathrm{s} \lesssim 300$. This behaviour is also corrected by the 
present fit modification.]

Let us remind, that the fit proposed in equation (4) of Paper III had 
the following form (note, that our $U$ is the energy per ion)
\begin{equation}
   \frac{U}{T} = \frac{3}{2}+ u_{\rm cl}(\Gamma) + u_{\rm q}~, 
\label{fit_en}
\end{equation}
where the first two terms on the right-hand side were the same as in 
the classic liquid \citep[][]{PC00}, while the new term $u_{\rm q}$ was
responsible for the ion quantum effects. To avoid a confusion, let us 
stress, that $u_{\rm cl}$ and $u_{\rm q}$  
are normalized to the temperature, whereas $\mathfrak u_\mathrm{ah}$ 
from Sec.\ \ref{fit_energy} is normalized to the typical Coulomb 
energy $Z_\mathrm{i}^2 e^2/a_\mathrm{i}$.

In this work, we suggest an updated expression for $u_{\rm q}$: 
\begin{eqnarray}
   u_{\rm q} &=& u_1+u_2+u_3~,
\label{uq123-} \\
   u_i &=& \frac{\theta_i}{1-\exp{(-\theta_i)}}-\frac{\theta_i}{2}-1~. 
\label{uq123}
\end{eqnarray}
In this case, $\theta_i = C_i \theta$, $i=1,2,3$, the 
quantities $C_1$, $C_2$, and $C_3$ are given by
\begin{equation}
       C_1 = \frac{P_1 r_\mathrm{s}}{P_3 + r_\mathrm{s}}~, ~~~
       C_2 = P_2~, ~~~
       C_3 = \sqrt{1 - C_1^2 - C_2^2}~,
\label{C-coef}
\end{equation}
and the coefficients $P_1$, $P_2$, and $P_3$ are summarized
in table \ref{P-coef}. By construction, equations 
(\ref{uq123-})--(\ref{C-coef}) reproduce the lowest-order term of 
the Wigner-Kirkwood expansion. 

It was observed in Paper III, on the basis of the data 
from Paper II (at $r_\mathrm{s} \geq 600$), that $u_{\rm q}$ depended
only on $\theta$ and not on temperature and density (or any other two 
parameters, such as $\Gamma$ and $r_\mathrm{s}$ or $\theta$ and 
$r_\mathrm{s}$) separately. This was not expected from the very 
beginning but stemmed 
from the fitting procedure. To emphasize this fact, the term was 
denoted $u_{\rm q}(\eta)$ in Paper III ($\eta$ of Paper III is 
our $\theta/\sqrt{3}$). Calculations at $r_\mathrm{s}=500$ 
revealed a subtle departure from this behavior. It was found, that 
$u_{\rm q}$ was better described by a function of $\theta$ 
[cf.\ equation (\ref{uq123})] with coefficients $C_1$ and $C_3$ 
depending explicitly on $r_\mathrm{s}$. This means that the 
quantum correction becomes a function of two variables: 
$u_{\rm q}(\theta,r_\mathrm{s})$. 

Accuracy of the new fit for 
$r_\mathrm{s} \gtrsim 1900$ is roughly the same as that of the 
Paper III fit, but, at $r_\mathrm{s} \lesssim 1900$, and especially 
at $r_\mathrm{s}=500$, the present formula fits the data better 
(cf.\ Fig.\ \ref{Fig_accuracy}). Additionally, we do not notice any 
unwanted effects, if the present fit is applied in the region 
$r_\mathrm{s} < 500$. More information on the properties of the new fit
is given in Appendix \ref{AppB}.

The new equation (\ref{uq123-}) for $u_{\rm q}$ results in 
a modification of quantum corrections to other ion thermodynamic 
functions and, in particular, to practical formulae (12)--(15) of 
Paper III describing them. Specifically, for the 
Helmholtz free energy (divided by $NT$), the last line of equation (12) 
of Paper III has to be replaced by $f_{\rm q} = f_1+f_2+f_3$, where 
\begin{equation}
    f_i = \ln{\left[\exp{(\theta_i)}-1\right]} 
    - \frac{\theta_i}{2} - \ln{\theta_i}~.
\label{f_i}
\end{equation}
For the isochoric specific heat, the last line of equation (13) of 
Paper III has to be replaced by $c_{\rm q} = c_1+c_2+c_3$,
where  
\begin{equation}
     c_i = \frac{\partial (T u_i)}{\partial T} = 
     u_i - \theta_i \frac{{\rm d} u_i}{{\rm d} \theta_i} =
     \frac{\theta_i^2 
    \exp{(\theta_i)}}{[\exp{(\theta_i)}-1]^2} - 1~. 
\label{c_i}
\end{equation}
Ion pressure becomes
\begin{equation}
      P = 
      P_{\rm id} \left( 1 + \frac{1}{3} u_{\rm cl} + \frac{1}{2} 
      u_{\rm q} - \frac{1}{3} \sum_{i=1}^3 D_i u_i \right)~,
\label{Pressi}
\end{equation}
where $P_{\rm id} = n_{\rm i} T$ is the ideal ion gas pressure, and
$D_i = {\rm d} \ln{C_i}/{\rm d} \ln{r_{\rm s}}$, so that
\begin{equation}
     D_1 = \frac{P_3}{P_3+r_{\rm s}}~, ~~~ D_2 = 0~, ~~~
     D_3 = - \frac{C_1^2}{C_3^2} \frac{P_3}{P_3+r_{\rm s}}~.             
\label{D_coef}
\end{equation}
For the temperature derivative of the ion pressure (divided by the ion 
density), the last line of equation (14) of Paper III has to be replaced
by
\begin{equation}
    \frac{1}{2} c_{\rm q} - \frac{1}{3} \sum_{i=1}^3 D_i c_i~.
\label{dPdTr}
\end{equation}
Finally, for the density derivative of the ion pressure (divided by 
$T$), the last line of equation (15) of Paper III has to be replaced by
\begin{equation}
        \frac{3}{4} u_{\rm q} - \frac{1}{4} c_{\rm q}
        - \frac{1}{2} \sum_{i=1}^3 D_i \left( u_i - \frac{1}{3} c_i 
        \right) 
        + \frac{1}{9} \sum_{i=1}^3 r_{\rm s} \frac{{\rm d} D_i}{{\rm d} 
        r_{\rm s}} u_i~. 
\label{dPdnr}
\end{equation}

\section{Thermodynamics across melting transition}
\label{Sec_Melt}
In this Section, we use the newly available thermodynamic information 
to obtain the phase diagram of the strongly-coupled ion plasma at 
astrophysically relevant densities and temperatures, neglecting 
ion-electron and higher-order electron contributions as described in 
the Introduction. The previous attempts at constructing the OCP phase 
diagram included the work of \citet{C93}, who used a quantum extension 
of the Lindemann criterion, and \citet{JC96}, who performed a PIMC 
study of the OCP. In both cases, the phase diagrams extended into 
the domain of very high densities, $r_\mathrm{s} \ll 500$, where ion OCP 
cannot 
exist due to rapid nuclear fusion reactions, which immediately change 
the composition and effectively increase $r_\mathrm{s}$ back to the 
range 
addressed in the present paper 
\citep[e.g.][and references therein]{B21}.

\begin{figure}                                           
\begin{center}                                              
\leavevmode                                                 
\includegraphics[width=\columnwidth]
{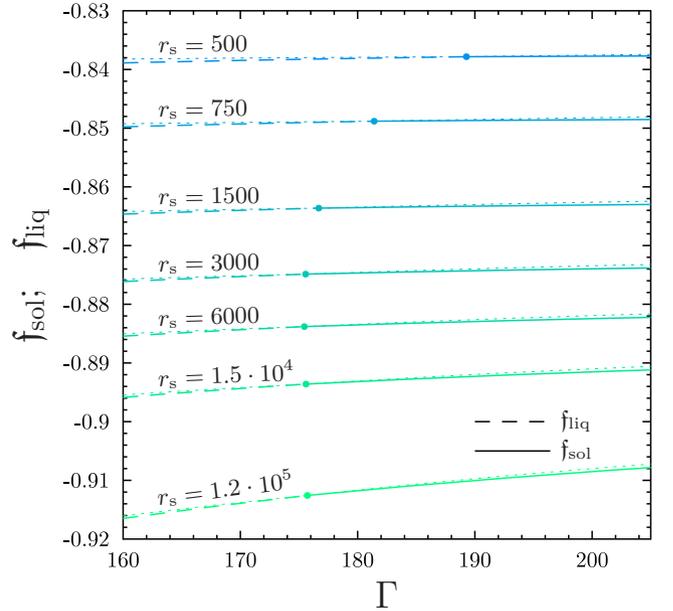} 
\end{center}                                                
\vspace{-0.4cm} 
\caption{Crystal (solid) and liquid (dashed) Helmholtz free energies
of the ion OCP 
[$\mathfrak f_{\rm sol, \,liq} \equiv a_\mathrm{i} 
F_{\rm solid, \,liquid}
/(Z^2 e^2)$] for a range of $r_\mathrm{s}$ values. The intersection 
point
of the curves, corresponding to the same $r_\mathrm{s}$, represents a 
phase 
transition. Dotted curves show a continuation of the liquid/solid free
energies into the supercooled/superheated regime.}                                             
\label{Fig_helm-eq}
\end{figure}
%

Let us begin by comparing the ion Helmholtz free energies of the 
liquid and solid phases. Their intersection point signals 
a melting/crystallization transition in the OCP, under assumption that
the ion number density is the same in both phases. This is illustrated 
in Fig.\ \ref{Fig_helm-eq}, where crystal and liquid ion Helmholtz 
free energies are displayed by solid and dashed lines, respectively, 
as functions of $\Gamma$ for several values of $r_\mathrm{s}$. 

\begin{figure}                                           
\begin{center}                                              
\leavevmode                                                 
\includegraphics[width=\columnwidth]
{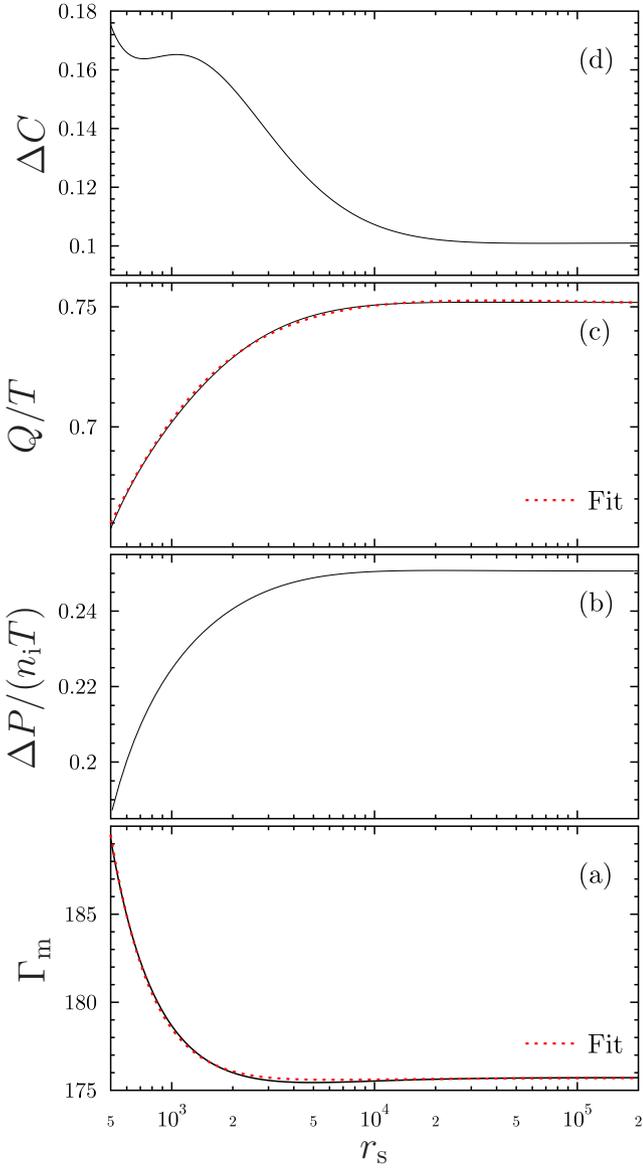} 
\end{center}                                                
\vspace{-0.4cm} 
\caption{Coulomb coupling strength at melting $\Gamma_{\rm m}$ 
(panel a), 
pressure jump at melting, specified as 
$F_{\rm liq} = F_{\rm sol}$, in units of $n_\mathrm{i} T$ (panel b),
latent heat $Q$ in units of $T$ (panel c), and 
specific heat jump at melting $\Delta C$ (panel d) vs. $r_\mathrm{s}$.}                                             
\label{Fig_GmQdC}
\end{figure}
%

The Coulomb coupling parameter at melting, $\Gamma_{\rm m}$, obtained 
in this way as a function of $r_\mathrm{s}$, is shown in panel (a) of 
Fig.\ \ref{Fig_GmQdC}. The dependence of $\Gamma_{\rm m}$ on 
$r_\mathrm{s}$ can be approximated by the following analytic expression 
\begin{equation}
     \Gamma_{\rm m} = 175.7-\frac{1300}{r_\mathrm{s}}
                       +\frac{4.1 \times 10^6}{r_\mathrm{s}^2}~.
\label{Fit_Gm}
\end{equation}
This formula is plotted by dots in panel (a) of 
Fig.\ \ref{Fig_GmQdC}.
It fits our numerical results in the range  
$500 \leq r_\mathrm{s} \leq 120000$.

\begin{figure}                                           
\begin{center}                                              
\leavevmode                                                 
\includegraphics[width=\columnwidth]
{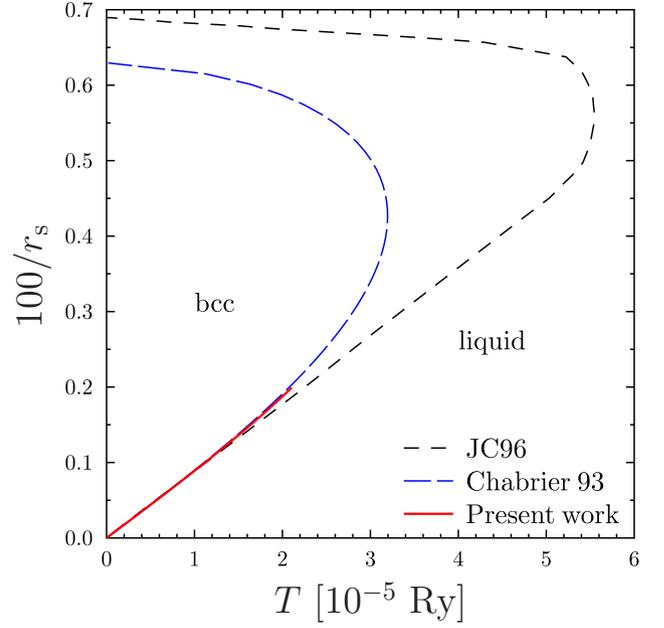} 
\end{center}                                                
\vspace{-0.4cm} 
\caption{Phase diagram of the OCP. Long and short dashes are melting
temperature predictions by \citet{C93} and \citet{JC96}, 
respectively. Solid line describes present results. 
}                                             
\label{phase-d}
\end{figure}

In Fig.\ \ref{phase-d}, we show the temperature-density plane as 
$100/r_\mathrm{s}$ versus $T/{\rm Ry}$ 
(${\rm Ry}= 0.5 m_\mathrm{i} Z_\mathrm{i}^4 e^4 /\hbar^2 = 
0.5 T \Gamma r_\mathrm{s}$ is the ion 
Rydberg). In this plot, long-dashed line is the crystallization curve 
predicted by \citet{C93}, short-dashed line is the prediction of 
\citet{JC96}, and solid line is the present prediction. Evidently, our 
prediction is closer to the result of \citet{C93}, but is not exactly 
the same.     

%

In reality though, a coexistence of phases is only possible at 
pressures 
equal on both sides, and the melting transition must be studied at a 
constant total pressure in both phases rather than at a fixed ion 
(and electron) density. The assumption of a fixed density implies an 
ion pressure jump between the phases, 
$\Delta P = P_{\rm liq} - P_{\rm sol}$. This jump can be 
easily 
calculated by combining the harmonic pressure from Paper I with 
present equations (\ref{Press}) and (\ref{Pressi}). It is shown in 
panel (b) of Fig.\ \ref{Fig_GmQdC}. 
In order to compensate the ion 
pressure jump, the density of the solid phase must be slightly higher, 
so that the electron pressure in the solid phase becomes higher than 
in the liquid. The required density increase is given by
\begin{equation}
    \delta n_\mathrm{i} = 
\Delta P \frac{\partial n_\mathrm{i}}{\partial P_{\rm sol}}~,  
\label{edni}
\end{equation}
where $P_{\rm sol}$ is the total solid pressure dominated by 
degenerate electrons. It is easy to show, that the density jump
will ensure the equality of the Gibbs free energies in both phases 
along with the pressure equality.  

\begin{figure}                                           
\begin{center}                                              
\leavevmode                                                 
\includegraphics[width=0.47\textwidth,bb=47 23 381 343, clip]
{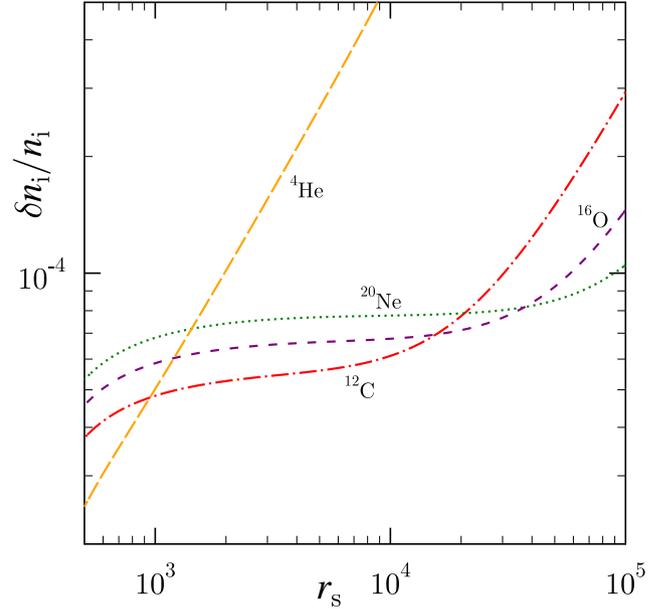} 
\end{center}                                                
\vspace{-0.4cm} 
\caption[]{Fractional ion density jump upon crystallization for nuclei 
species typical of matter in a WD core.}                                             
\label{dn}
\end{figure}

Due to the density increment, the resulting $\Gamma_{\rm m}$ at melting
becomes different in the liquid and solid phases, but this
difference as well as the difference with the original, fixed density,
$\Gamma_{\rm m}$ is extremely small. If we were to plot these quantities
versus $r_\mathrm{s}$ in Fig.\ \ref{Fig_GmQdC}, the three curves would 
merge.
Indeed, in Fig.\ \ref{dn}, we show the fractional density jump between 
the liquid 
and solid phases as a function of $r_\mathrm{s}$. 
To calculate this quantity, we employed exact formulae for ion 
thermodynamic functions and the standard thermodynamics of the 
degenerate
electron gas at $T=0$ \citep[e.g.][]{HPY07}. The density jump depends 
on the assumed ion species, because the functional form of the density 
dependence of the electron pressure is sensitive to the electron 
relativity degree.

%

If the phase transition is found from the equality of the Helmholtz 
free energies at fixed density (as in the beginning of this Section), 
one sees that the energy difference between the phases, 
$Q \equiv U_{\rm liq} - U_{\rm sol} > 0$, is equal to 
$T \Delta S$. 
In this 
case, $\Delta S$ is the ion entropy difference. Introducing 
the 
density 
jump, equation (\ref{edni}), and imposing the equality of the Gibbs free 
energies,
one sees, that the same $T \Delta S$ (ion quantities 
essentially do 
not change as a consequence of this procedure) becomes equal to the
enthalpy difference between the phases also known as the latent heat 
of crystallization. This quantity plays an important role in 
astrophysics of white dwarfs. It is 
released as the matter in a WD core crystallizes, which 
delays cooling of these stars and affects various age estimates 
\citep[e.g.][and references therein]{Tetal19}. 
According to our calculations, the latent heat is a slowly varying 
function of $r_\mathrm{s}$,
which is shown in panel (c) of 
Fig.\ \ref{Fig_GmQdC}. This dependence can be approximated by an 
analytic expression valid for $500 \leq r_\mathrm{s} \leq 120000$, as 
follows:
\begin{equation}
     \frac{Q}{T} = 
         0.75 \left( 1 + \frac{1.4}{\sqrt{r_\mathrm{s}}}
         -\frac{145}{r_\mathrm{s}}
         +\frac{1200}{r_\mathrm{s}\sqrt{r_\mathrm{s}}} \right)~.
\label{Fit_q}
\end{equation}
This formula is plotted by dots in panel (c) of 
Fig.\ \ref{Fig_GmQdC}.

Note, that at higher $r_\mathrm{s}$, where ions are almost 
classic, our calculations indicate that 
$\Delta P \approx n_\mathrm{i} Q/3$ in agreement with theory 
\citep*[][]{HTV77}. Let us also observe that $\Delta S = Q/T$ 
is the jump of {\it excess entropy} at melting.  

\begin{figure}                                           
\begin{center}                                              
\leavevmode                                                 
\includegraphics[width=\columnwidth]
{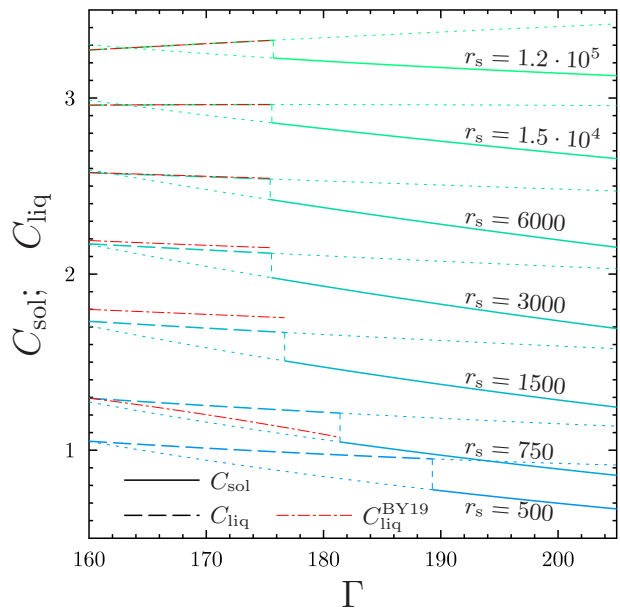}  
\end{center}                                                
\vspace{-0.4cm} 
\caption{Specific heat of the OCP vs. $\Gamma$ for several values of 
$r_\mathrm{s}$. Discontinuity indicates melting transition. Solid 
curves, 
$C_{\rm sol}$, and dashed curves, $C_{\rm liq}$, represent the crystal 
and liquid phases, respectively. Dotted curves demonstrate the 
specific heat of a superheated crystal (at lower $\Gamma$) and 
a supercooled liquid (at higher $\Gamma$). Thin dot-dashed curves 
show the specific heat in the liquid based on the Paper III fit. }                                             
\label{Fig_specmelt}
\end{figure}

Another remarkable quantity is the specific heat jump between the solid
and liquid phases. In Fig.\ \ref{Fig_specmelt}, we show the ion specific 
heat of the OCP across the phase transition. In this case, we combine 
updated thermodynamics of the liquid from Sec.\ \ref{liq_upd}, Paper I 
for the harmonic term in 
the crystal, and the results of Sec.\ \ref{fit_thermodyn} for the 
anharmonic contribution. 
Different curves correspond to different $r_\mathrm{s}$ values. The 
presence of 
the jump is evident and so is its $r_\mathrm{s}$ dependence. The 
dependence of 
the specific heat jump at melting on $r_\mathrm{s}$ is shown in panel 
(d) of Fig.\ \ref{Fig_GmQdC}. 

Thin dot-dashed curves in Fig.\ \ref{Fig_specmelt} display the specific 
heat in the liquid phase obtained from the Paper III fit. It is seen 
to be sufficiently accurate at lower $\Gamma$ or higher $r_\mathrm{s}$, 
but gradually worsens, as one approaches the phase transition in the 
deep quantum regime.

%

\section{Astrophysical applications}
\label{Sec_apply}
As discussed in the Introduction, one-component plasma model is relevant
for astrophysics of compact degenerate stars, white dwarfs and neutron
stars. In these objects, matter at densities above 
$\sim 10-1000$ g cm$^{-3}$ (depending on the chemical composition) is
fully pressure-ionized. The phase, containing bare atomic nuclei and
nearly rigid, degenerate electron gas, extends all the way down to the
center of a star, in the case of white dwarfs, or to the outer boundary
of the inner crust (at $\rho_{\rm d} \approx 4.3 \times 10^{11}$ 
g cm$^{-3}$), in the case of neutron stars.

The chemical elements present in such matter may vary from the lighter
H, He, and C to the heavier Ne, Fe, and many other. As we saw in 
Secs.\ \ref{fit_energy}, \ref{fit_thermodyn}, and \ref{liq_upd}, ion 
thermodynamics is parameterized by just two
dimensionless quantities, $\Gamma$ and $r_{\rm s}$. Consequently, our
fitting formulae enable one to determine all the necessary thermodynamic
functions in the entire temperature and density range of astrophysical 
interest for any nuclear species.

Let us illustrate the results of the preceding sections by considering
a few fiducial astrophysical examples. We shall focus on the total heat 
capacity (ion plus electron) of fully ionized matter. The ion 
contribution to it is probably the most practically important ion 
thermodynamic function, as it dominates over the electron contribution 
and thus has a significant impact on the thermal evolution of a star.

\begin{figure}                                           
\begin{center}                                              
\leavevmode                                                 
\includegraphics[width=\columnwidth]
{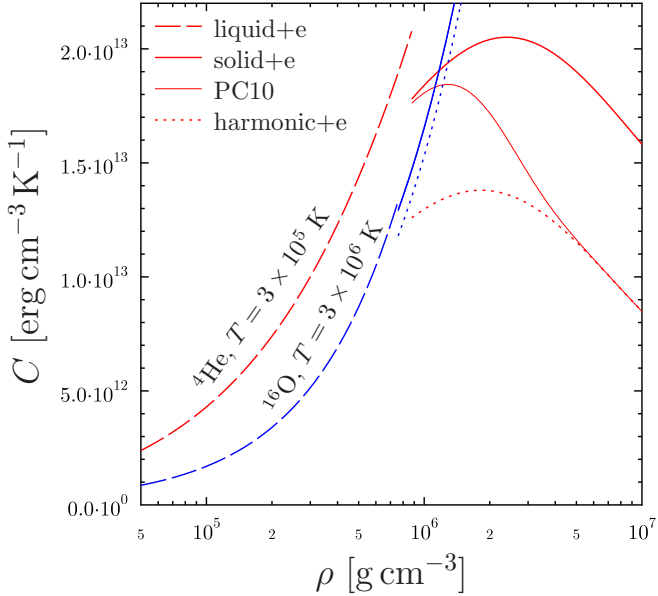}  
\end{center}                                                
\vspace{-0.4cm} 
\caption{Total (ion plus electron) heat capacity  of dense matter 
composed of helium at $T=3 \times 10^5$ K or oxygen at $T=3 \times 10^6$ K 
vs. mass density. Thick long-dashed and 
solid curves correspond to the liquid and crystal phases, respectively.
Dotted lines include only ion harmonic and electron contributions. 
Thin solid line (merging with the thick solid line in the 
case of oxygen) is the parameterization of \citet{PC10}.}                                             
\label{Fig_hc_HeO}
\end{figure}

In Fig.\ \ref{Fig_hc_HeO}, we plot, in physical units, the total heat 
capacity of dense matter composed of helium at $T=3 \times 10^5$ K or 
oxygen at $T=3 \times 10^6$ K as a function of mass density. Thick 
long-dashed lines describe the liquid phase. Thick solid curves is 
the total heat capacity of the solid phase. Dots do not include the 
ion anharmonic contribution, whereas thin solid curves demonstrate the 
total solid-phase heat capacity as parameterized by \citet{PC10} but
without the ion-electron contribution 
associated with electron screening.

%

\begin{figure}                                           
\begin{center}                                              
\leavevmode                                                 
\includegraphics[width=\columnwidth]
{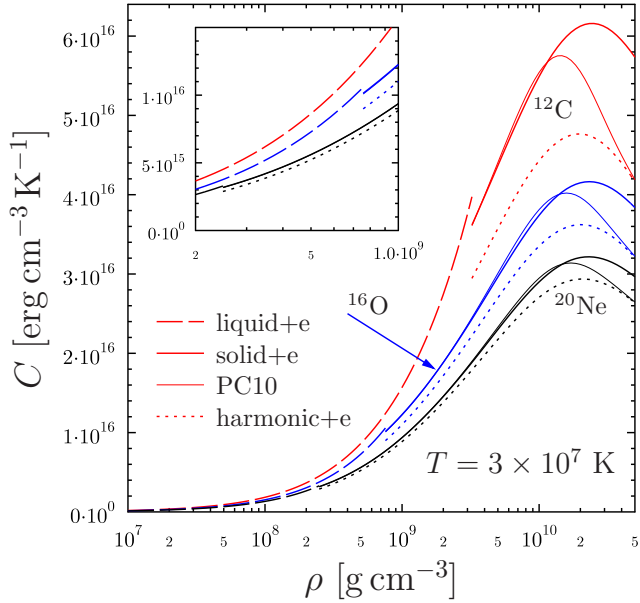}  
\end{center}                                                
\vspace{-0.4cm} 
\caption{Same as in Fig.\ \ref{Fig_hc_HeO} but for carbon, oxygen, 
and neon plasma at $T=3 \times 10^7$ K.}                                             
\label{Fig_hc_CONe}
\end{figure}
%

In Fig.\ \ref{Fig_hc_CONe}, the total heat capacity of carbon, oxygen, 
and neon plasma at $T=3 \times 10^7$ K is shown as a function of 
density. As the matter becomes hotter and the nuclei heavier, the 
system becomes more classic and the quantum effects are less pronounced.
Nevertheless, one still observes a discrepancy with the parameterization
of \citet{PC10}, which can over- or underestimate the actual heat 
capacity.

In real systems, such as matter in a WD core or an NS crust, one expects 
to find a mixture of several ion species rather  
than the OCP. In order to describe such systems from first principles,
one has to introduce new free parameters (one per each additional 
species) and perform numerous extra simulations. Such a program for 
quantum plasmas requires enormous computer resources and has not been 
undertaken yet for exhaustive parameter space (see, however, 
\citealt{RR16}).
For the crystallized phase, there is an additional complication 
associated with the fact that the structure of a multi-component 
crystal is still not fully understood even in the classic case 
(see \citealt{Caplan20} for recent progress).

\section{Conclusion} 
\label{Sec_conclus}
We have performed detailed first-principle path-integral Monte Carlo 
simulations of a bcc crystal composed of ions of a single sort and 
rigid uniform charge-compensating electron background (one-component
plasma crystal). The study was focused on the density range 
$500 \leq r_\mathrm{s} \leq 120000$, covering fully-ionized layers in 
the 
interior of white dwarfs and neutron stars, at temperatures, spanning 
the range from the crystallization line down to $T_{\rm p}/30$. 
Ion quantum effects are not negligible in a large fraction of this 
physical parameter domain.  

The crystal anharmonic energy has been extracted from the simulation 
results by subtracting the harmonic energy of a finite lattice. The 
anharmonic energy has been fitted by a convenient analytic formula 
(\ref{U_anh}), which incorporated available classic and quantum 
asymptotes. 

The 
fitting formula
has been integrated and differentiated, and explicit expressions for
anharmonic contributions to the crystal Helmholtz free energy, entropy,
specific heat, and pressure are given.
These are equations (\ref{F_anh}), (\ref{S_anh}), (\ref{C_anh}), 
and (\ref{Press}), which should be 
included in evolutionary codes to ensure accurate treatment 
of quantum ion thermodynamics of a crystal. These formulae cover the 
entire crystallized region of WD or NS interior, where they supercede 
any previously available results. 

Naturally, the effect on the results 
of the evolutionary codes is not expected to be drastic, because the 
crystal thermodynamics is dominated by the well-known harmonic and 
electron contributions. Nevertheless, certain thermal properties, most  
importantly the specific heat, will receive a boost of up to 50\%, 
depending on density and temperature, which is potentially noticeable.

In parallel to this work, the previously available analytic description
of the liquid-phase thermodynamics of the same system has been
updated to include new PIMC data at $r_\mathrm{s}=500$ and in the 
vicinity of the crystallization curve. We recommend replacing 
equations (6) and (9) of Paper III by the present
equations (\ref{uq123-}) and (\ref{Pressi}), combining equations 
(12)--(15) of Paper III
with the present equations (\ref{f_i}), (\ref{c_i}), (\ref{dPdTr}), 
and (\ref{dPdnr}), and incorporating them into evolutionary codes to 
describe 
thermodynamics of a quantum ion liquid. These formulae cover the 
entire liquid region of WD core or NS crust, where they supercede 
any previously available results.

By combining the 
liquid and
solid thermodynamics, we have analyzed in detail melting properties of 
the OCP with account of ion quantum effects. In particular, we 
have obtained from the first principles the density dependence of: 
({\it i})  the Coulomb coupling strength at melting, $\Gamma_{\rm m}$; 
({\it ii}) the energy jump, which is equal to the latent heat of 
crystallization, $Q$; 
({\it iii}) the pressure jump, accompanying the phase transition, if 
it is determined by equating the liquid and solid Helmholtz free 
energies; 
({\it iv}) the respective density jump, which appears in the 
proper formalism based on the Gibbs free energy and accompanies 
crystallization in real systems;  
and, finally, ({\it v}) the specific heat jump. All these functions 
are displayed in Figs.\ \ref{Fig_GmQdC} and \ref{dn}. The dependences 
of $\Gamma_{\rm m}$ and $Q$ on density have been fitted by simple 
analytic formulae (\ref{Fit_Gm}) and (\ref{Fit_q}). 

These results are crucial for reliable modelling 
of thermal evolution and asteroseismology of compact degenerate stars.

\section*{Acknowledgments}
We would like to acknowledge the use of the {\it Maxima} package for 
symbolic computations. This work was partially supported by the 
Russian Science Foundation grant 19-12-00133.

\section*{Data Availability}

The data underlying this article will be shared on reasonable 
request to the corresponding author.



\appendix

\section{Harmonic lattice in PIMC simulations}
\label{AppA}
Within the harmonic lattice approximation, the interionic interaction 
potential is expanded up to the second order in ion displacements from 
the lattice sites. Since the lattice sites correspond to 
equilibrium positions, the linear terms are absent, and the dynamics 
of the ion system can be conveniently described by a set of 
independent collective coordinates (phonon modes). Each mode 
behaves as a harmonic oscillator with the respective frequency.

In comparison with the well-known thermodynamic limit (e.g. Paper I), 
PIMC simulations of a crystal include a finite number of ions $N$ in 
a box with periodic boundary conditions. Moreover, quantum effects are 
modelled by the PIMC approach, i.e. by using the primitive 
approximation with a finite number $M$ of imaginary time slices. Thus,
to subtract the harmonic energy from the PIMC data, the harmonic lattice 
formalism must be developed for the same finite $N$. Then, the 
ion motion will be still described by a set of independent collective 
modes, but only those phonons will be present, whose wavevectors are 
consistent with the periodic boundary conditions. Furthermore, each 
finite-$N$ phonon mode must be treated in the primitive approximation 
with $M$ imaginary time slices. As we demonstrate in the next two 
subsections, both types of finite-size effects can be accounted for 
analytically.

\subsection{Harmonic oscillator in PIMC} 
\label{Sec_HarmOsc_PIMC}
Accurate thermodynamics of a harmonic oscillator (ho) with a frequency 
$\omega$ is well known (e.g. \citealt*{DW20,VVNG97}, where the relation 
to 
the PIMC approach is explicitly analyzed). In particular, 
the partition function $Z_\mathrm{ho}=\exp(z/2)/[\exp(z)-1]$ and
the energy $U_\mathrm{ho}=U_\mathrm{ho}^0+\hbar \omega/[\exp(z)-1]$. 
In this case, $z=\hbar \omega/T$ and $U_\mathrm{ho}^0=\hbar \omega/2$ 
is the zero-point energy.

In a PIMC simulation, quantum harmonic oscillator is modelled by a 
classic ring polymer with $M$ beads in an effective potential
\begin{equation}
H=\frac{\mu\omega^2}{2 M} \sum_i r_i^2+\frac{\mu M T^2}{2\hbar^2}
\sum_i (r_i-r_{i+1})^2,
\end{equation}
where $\mu$ is the mass and $r_i$ is a (one-dimensional) position 
of the $i$-th bead ($r_M=r_0$ due to the ring condition). The second 
term is related to the kinetic energy. 
Let us consider thermodynamics of this system following \cite{VVL03}.  
The partition function of this polymer can be written in an explicit 
form 
\begin{eqnarray}
Z_{\mathrm{ho},M}&=&\left(\frac{M }{2\pi z}\right)^{M/2}\,
\int \exp\left[-\left(\frac{M}{z}+\frac{z}{2M}\right)\sum_i x_i^2
\right. 
\nonumber \\
&&\left. +\frac{M}{z}\sum_i x_i x_{i+1}\right]
\mathrm d x_1\ldots\mathrm d x_M,
\end{eqnarray}
where coordinates $x_i$ are measured in natural units 
$\sqrt{\hbar/(\mu\omega)}$. The exponent in square brackets is a 
quadratic form over $x_i$. Changing variables to the main axes of this 
form, \cite{VVL03} arrive at 
$Z_{\mathrm{ho},M}=(\det \mathbf{A}_M)^{-1/2}$, 
with the matrix
\begin{equation}
\mathbf{A}_M=
\begin{pmatrix}
2\,C &-1 &0 &\cdots &0& -1\\
-1 & 2\,C& -1& 0& \cdots& 0\\
0& -1 & 2\,C& -1& \cdots& 0\\
\vdots & & &  \ddots & & \vdots\\
0& \cdots & 0& -1& 2\,C & -1\\
-1& 0& \cdots & 0 &-1& 2\,C \\
\end{pmatrix}
\end{equation}
and $C=1+(z/M)^2/2$. \cite{VVL03} also obtained several useful recurrent 
relations to determine $\det \mathbf{A}_M$. Using mathematical 
induction, 
we write down $\det \mathbf{A}_M$ in an explicit form:
\begin{equation}
\det \bm A_M=\sum_{i=0}^{M-1} \mathfrak{A}_M^i 
\left(\frac{z}{M}\right)^{2(i+1)},
\end{equation}
where
\begin{equation}
\mathfrak{A}_M^i=2\frac{M^2\,(M^2-1^2)\,(M^2-2^2)
	\cdots (M^2-i^2)}{(2i+2)!}.
\end{equation}

Knowing the partition function, it is easy to calculate the energy of 
the harmonic oscillator for $M$ beads
\begin{equation}
U_{\mathrm{ho},M}=-\frac{\partial 
	\ln{Z_{\mathrm{ho},M}}}{\partial (1/T)}
=\frac{\hbar \omega}{2}\frac{\mathrm d \det \mathbf{A}_M}{\mathrm d z}
\frac{1}{\det \mathbf{A}_M}.
\label{Uharmosc_PIMC}
\end{equation}
%

\begin{figure}                                           
\begin{center}                                              
\leavevmode                                                 
\includegraphics[width=\columnwidth]
{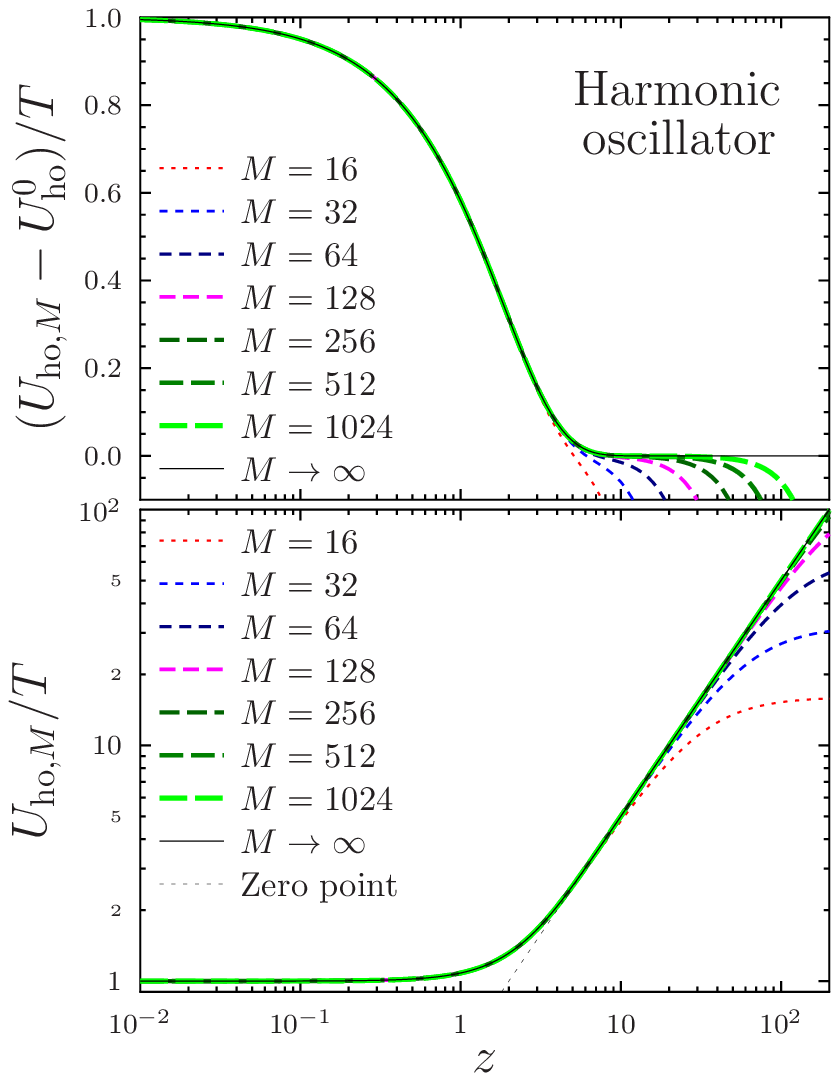} 
\end{center}                                                
\vspace{-0.4cm} 
\caption{Energy of the harmonic oscillator $U_{\mathrm{ho},M}/T$ 
(bottom panel) and its ``thermal energy'' 
$(U_{\mathrm{ho},M}-U_\mathrm{ho}^0)/T$ (top panel) vs. $z$
as obtained in a PIMC simulation with  
$M=16$, $32$, $64$, $128$, $256$, $512$, and $1024$ beads.}                                             
\label{Fig_U_PIMC}
\end{figure}
%

The energy of the harmonic oscillator, as predicted by the PIMC 
simulation, is shown in 
Fig.\ \ref{Fig_U_PIMC} for several values of  $M$. Solid line 
represents the accurate quantum limit, corresponding to 
$M\rightarrow \infty$. Upper panel represents the ``thermal energy'' 
$(U_\mathrm{ho}-U_\mathrm{ho}^0)/T$.
One can see, that the PIMC simulation, for a fixed $M$, can reproduce 
the accurate quantum energy only up to some value $z$. At higher $z$, 
it strongly underestimates the energy (in particular, it does not 
reproduce the zero-point energy). It is worth noting, that for 
$zM\lesssim 1$, the difference between the PIMC energy and the exact 
quantum energy is well fitted (with an error of less than 
$0.01\hbar \omega$) by
\begin{equation}
U_{\mathrm{ho},M}-U_\mathrm{ho}
=-\frac{\hbar\omega}{16}\frac{z^2}{M^2}.
\label{ho_correction_to_PIMC}
\end{equation}

\subsection{Finite-size effects for harmonic lattice}
\label{Sec_HL_PIMC}
We consider a bcc lattice and a cubic simulation box, thus, the number 
of ions in simulation
$N=2 N_{L}^3$, where $N_{L}$ is the number of cubic cells in each 
direction of the simulation box.
Obviously, the box size is $L=N_L a_1$, where $a_1$ is the size of the 
bcc cubic cell. Thanks to the symmetry of the bcc lattice,
it suffices to consider phonons only from the so-called irreducible 
part of the Brillouin zone, which is $1/48$-th of the whole 
Brillouin zone (e.g. \citealt{AG81}). Their wavevectors can be 
specified as:
$\bm k=(2\pi i_x/L,2\pi i_y/L,2\pi i_z/L)$, where $i_x=0,\ldots,N_l$, 
$i_y=0,\ldots, \min(i_x,N_L-i_x)$, 
$i_z=0, \ldots, i_y$.

\begin{figure}                                           
\begin{center}                                              
\leavevmode                                                 
\includegraphics[width=\columnwidth]{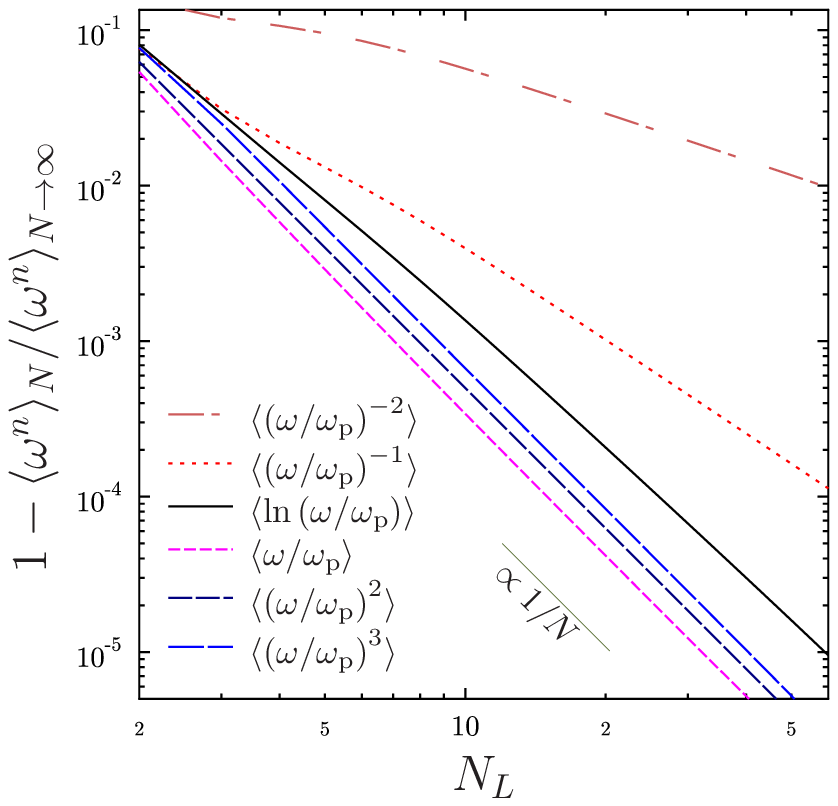} 
\end{center}                                                
\vspace{-0.4cm} 
\caption{The relative finite-size correction for several phonon 
frequency moments as a function of $N_L$. Thin solid line indicates 
a $\propto 1/N=1/(2\,N_L^3)$ dependence.
}                                             
\label{Fig_FreqMom}
\end{figure}
%

There are three modes for each wavevector, frequencies of these modes 
$\omega_j$ ($j=1,\ 2,\ 3$) are determined by the dynamic matrix of the 
bcc lattice for the OCP 
(see, e.g. Paper I for an explicit form).
In particular, one can calculate phonon frequency moments according to
\begin{equation}
\langle \omega^n \rangle_N =\frac{1}{3N}
    \sum_{i_x,i_y,i_z} w(i_x,i_y,i_z) 
    \sum_j \omega_j^n~.
\end{equation}
In this case, $w(i_x,i_y,i_z)$ is the weight factor, determined by the 
symmetry (\citealt{W72_thermodynamics,AG81}). 
Due to finite $N$, the resulting phonon moments differ from the well 
known values in thermodynamic limit (see, e.g. Paper I). The relative 
difference is shown in Fig.\ \ref{Fig_FreqMom}. Note, that  
contribution of $k=0$ phonons was taken to be zero for all phonon 
moments to avoid divergence of some of them (see also below). It is 
the only reason there is a finite-size effect for the second moment 
$\langle \omega^2 / \omega_{\rm p}^2 \rangle_N$. In fact, thanks to 
the Kohn sum rule (\citealt{BP55,Brout59}), 
$\sum_j \omega_j^2=\omega_{\rm p}^2$ for any phonon wavevector. Thus,  
$\langle \omega^2 / \omega_{\rm p}^2\rangle_{N}\equiv 1/3-1/(3N)$, 
where the last term is associated with the zero contribution of 
$k=0$ phonons. 
As is obvious from Fig.\ \ref{Fig_FreqMom}, the relative finite-size 
correction for moments with a positive $n$ follows $\propto 1/N$ 
trend, but it is not so for $n<0$ moments and for the 
$\langle \ln{(\omega/\omega_{\rm p})}\rangle_N$ moment. This is 
because the dominating 
contribution to the latter moments comes from the vicinity of the 
Brillouin zone center, and thus strongly depends on the location of 
phonon wavevectors, available at finite $N$.

\begin{figure}                                           
\begin{center}                                              
\leavevmode                                                 
\includegraphics[width=\columnwidth]{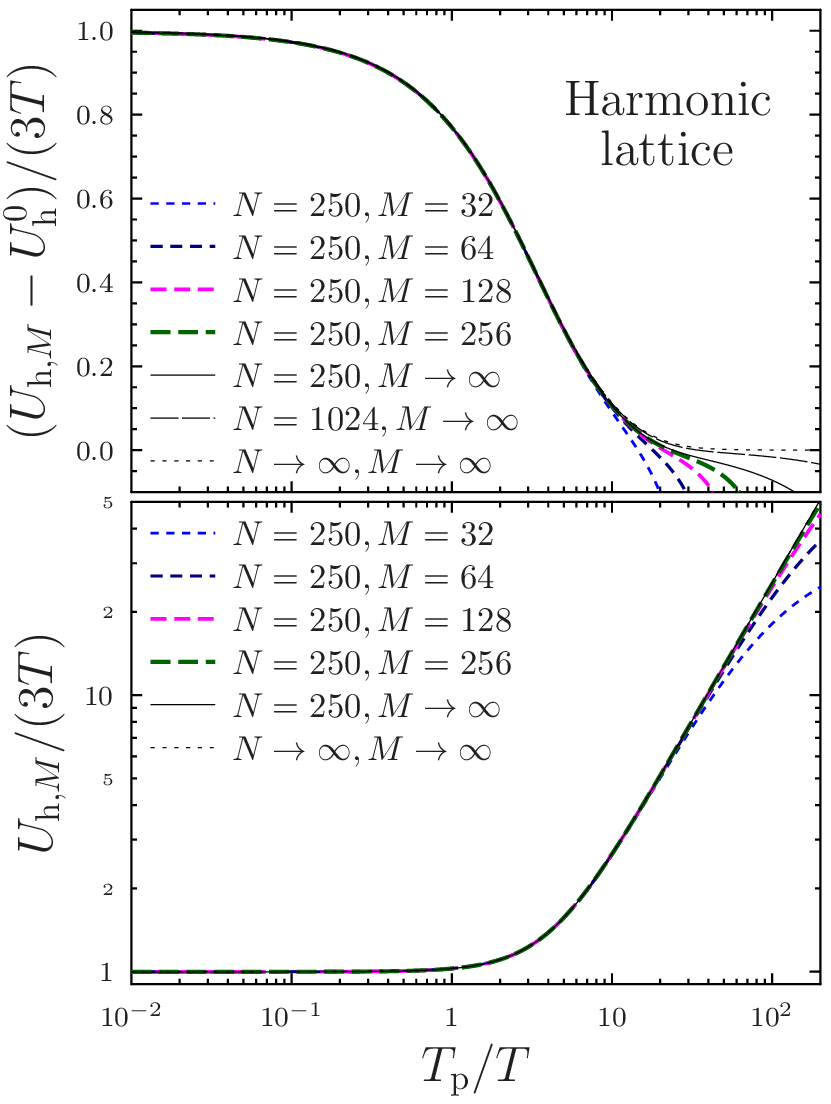} 
\end{center}                                                
\vspace{-0.4cm} 
\caption{Harmonic lattice energy 
$U_{\mathrm{h},M}/(3T)$ (bottom 
panel) and ``thermal energy'' $(U_{\mathrm{h},M}-U_\mathrm{h}^0)/(3T)$ 
(top panel) as functions of $T_\mathrm p/T$ for a PIMC harmonic lattice 
simulation with $N=250$ and $M=32$, $64$, $128$, or $256$ beads. 
Thin dotted line corresponds to the accurate quantum thermodynamic 
limit, while thin solid and long-dashed lines provide an accurate 
quantum description ($M\rightarrow \infty$) of finite 
systems with 250 and 1024 particles, respectively.}                                             
\label{Fig_UHL_PIMC}
\end{figure}
%

The harmonic energy per one ion is calculated as 
\begin{equation}
U_{\mathrm{h},M}=\frac{1}{N}\sum_{i_x,i_y,i_z} w(i_x,i_y,i_z) \sum_j
U_{\mathrm{ho},M}(\omega_j,T).
\label{UhMN}
\end{equation}
$U_{\mathrm{ho},M}(\omega_j,T)$ is the energy given by 
equation (\ref{Uharmosc_PIMC}).
Note, however, that modes with ${\bm k}=0$ require special attention. 
They correspond to a motion of all ions together, i.e. a motion of the 
center of mass. As long as we consider only interionic interactions, 
it is a free motion not constrained by any forces. 
Obviously, the contribution of ${\bm k}=0$ modes must be equal to 
$3T/2$. 

The resulting harmonic lattice energy as a function of 
$T_\mathrm p/T$ (cf.\ Sec.\ \ref{fit_energy}) is shown in the bottom 
panel 
of  Fig.\ \ref{Fig_UHL_PIMC}. The dotted line 
represents the energy 
in the thermodynamic limit (Paper I);
in the scale of the bottom panel, it coincides with the solid line, 
which represents accurate 
quantum description for $N=250$.
Similar to the harmonic 
oscillator (Fig. \ref{Fig_U_PIMC}), PIMC underestimates the energy at 
low $T$ due to a finite number of beads.

To demonstrate the importance of the finite-size effects in the strongly 
quantum regime ($T\ll T_\mathrm p$), in the top panel of 
Fig.\ \ref{Fig_UHL_PIMC}, we plot 
$(U_{\mathrm{h},M}-U_{\mathrm{h}}^0)/3T$ as a function of 
$T_\mathrm p/T$. In this case, 
$U_{\mathrm{h}}^0/N=\hbar \omega_\mathrm{p}\langle \omega/
\omega_\mathrm{p}\rangle/2$ is 
the zero-point energy calculated in the thermodynamic limit (the 
first frequency moment 
$\langle \omega/\omega_\mathrm{p}\rangle=0.5113875$, Paper I).
As it should be, in the thermodynamic limit, this difference represents 
the thermal energy and it is positive (see the dotted line).  However, 
it is not the case for the finite-$N$ simulation, even if the quantum 
effects are included accurately ($M\rightarrow \infty$, see thin solid 
line). This is because the first frequency moment for $N=250$ 
($N_L=5$) is a bit lower: 
$\langle \omega/\omega_\mathrm{p}\rangle_{250}=0.5099073$ (as 
expected, the 
correction is of the order of $1/N$, see Fig.\ \ref{Fig_FreqMom}).
Usage of a finite number of beads $M$ in a PIMC simulation makes 
agreement with the thermodynamic limit even 
worse. That is why we use $U_{\mathrm{h},M}$ calculated from 
equation (\ref{UhMN}) to extract the anharmonic energy from our 
simulations. In this 
way, we avoid an impact of the finite-size effects in the main, 
harmonic, energy on the subdominant anharmonic energy.

\section{Extrapolation properties of the new liquid fit}
\label{AppB}
\begin{figure}                                           
\begin{center}                                              
\leavevmode                                                 
\includegraphics[width=\columnwidth]
{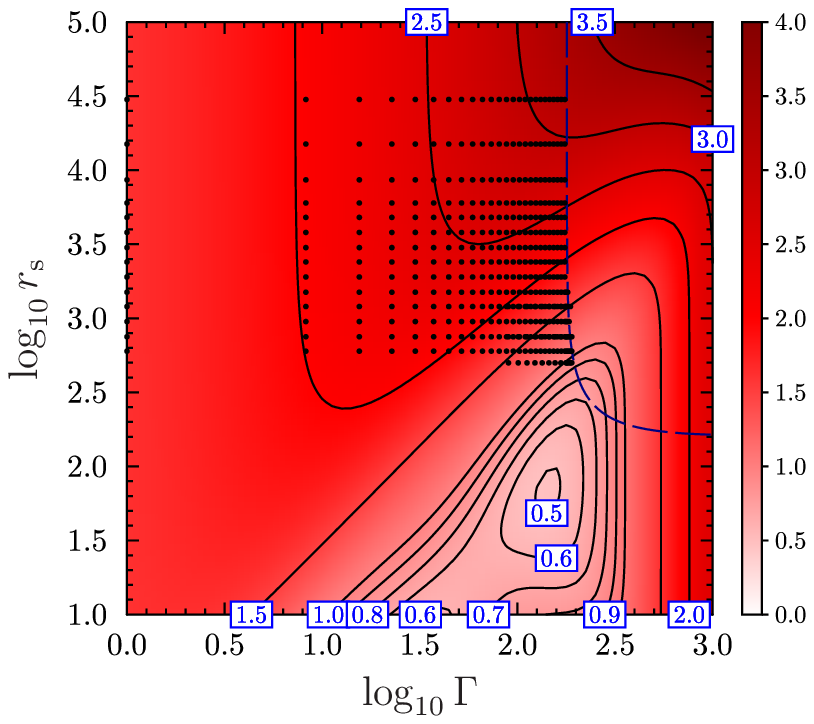} 
\end{center}                                                
\vspace{-0.4cm} 
\caption{Color map of the ion specific heat derived from the 
new liquid-phase fit. Dashed line represents 
the melting curve, dots indicate PIMC points in the liquid, on which 
the fit is based.}                                             
\label{Fig_CvMap}
\end{figure}

In Fig.\ \ref{Fig_CvMap}, we show a color map of the specific heat 
derived from the new fit of the liquid phase energy, equations 
(\ref{fit_en}) and (\ref{uq123-}), in a very broad 
domain of $\Gamma$ and $r_\mathrm{s}$, covering the solid phase as well 
as the extremely dense region $r_\mathrm{s} \ll 500$. The 
specific heat is explicitly positive in the extrapolated region. A 
minimum at $\Gamma \approx 140$ and $r_\mathrm{s}\approx 70$ is likely 
an artefact of our fit. Note however, that this artefact 
should not affect any astrophysical applications, because it is located 
far outside the region of astrophysical interest.

%

\label{lastpage}
\end{document}